\documentclass[10pt,twocolumn,preprintnumbers,amsmath,amssymb,nofootinbib
,superscriptaddress]{revtex4-1}
\usepackage{graphicx,longtable,mathrsfs,color,array}
\usepackage[usenames,dvipsnames]{xcolor} 
\usepackage{amssymb,amsmath,mathtools,mathrsfs,slashed} 
\usepackage{epsfig,subfigure,placeins,float} 
\usepackage{booktabs,longtable,ctable,multirow} 
\usepackage{exscale,relsize} 
\usepackage[normalem]{ulem} 
\usepackage{enumerate}
\usepackage{color}
\usepackage{hyperref}
\usepackage{comment}
\allowdisplaybreaks[1]

\newcommand{\mf}[1]{\textcolor{cyan}{\textbf{[MF: #1]}}}

\newcommand{\xGW}{x_\mathrm{\rm GW}}
\newcommand{\xEM}{x_\mathrm{EM}}
\newcommand{\DGW}{d_\mathrm{\rm GW}}

\begin{document}
\title{Standard sirens with a running Planck mass}
\author{Macarena Lagos}
\email{mlagos@kicp.uchicago.edu}
\affiliation{Kavli Institute for Cosmological Physics, The University of Chicago, Chicago, IL 60637, USA}
\author{Maya Fishbach}
\email{mfishbach@uchicago.edu}
\affiliation{Department of Astronomy and Astrophysics, University of Chicago, Chicago, IL 60637, USA}
\author{Philippe Landry}
\email{landryp@uchicago.edu}
\affiliation{Kavli Institute for Cosmological Physics, The University of Chicago, Chicago, IL 60637, USA}
\affiliation{Enrico Fermi Institute, University of Chicago, Chicago, IL 60637, USA}
\author{Daniel E. Holz}
\email{holz@uchicago.edu}
\affiliation{Kavli Institute for Cosmological Physics, The University of Chicago, Chicago, IL 60637, USA}
\affiliation{Department of Astronomy and Astrophysics, University of Chicago, Chicago, IL 60637, USA}
\affiliation{Enrico Fermi Institute, University of Chicago, Chicago, IL 60637, USA}
\affiliation{Department of Physics, University of Chicago, Chicago, IL 60637, USA}
\date{Received \today; published -- 00, 0000}

\begin{abstract}
We consider the effect of a time-varying Planck mass on the propagation of gravitational waves (GWs).
A running Planck mass arises naturally in several modified gravity theories, and here we focus on those that carry an additional dark energy field responsible for the late-time accelerated expansion of the universe, yet---like general relativity (GR)---propagate only two GW polarizations, both traveling at the speed of light. Because a time-varying Planck mass affects the amplitude of the GWs and therefore the inferred distance to the source, standard siren measurements of $H_0$ are degenerate with the parameter $c_M$ characterizing the time-varying Planck mass, where $c_M=0$ corresponds to GR with a constant Planck mass.  
The effect of non-zero $c_M$ will have a noticeable impact on GWs emitted by binary neutron stars (BNSs) at the sensitivities and distances observable by ground-based GW detectors such as advanced LIGO and A+, implying that standard siren measurements can provide joint constraints on $H_0$ and $c_M$.  Assuming a $\Lambda$CDM evolution of the universe and taking \textit{Planck}'s measurement of $H_0$ as a prior, we find that GW170817 constrains $c_M = -9^{+21}_{-28}$ ($68.3\%$ credibility).
We also discuss forecasts, finding that if we assume $H_0$ is known independently (e.g.~from the cosmic microwave background), then 100 BNS events detected by advanced LIGO can constrain $c_M$ to within $\pm0.9$. This is comparable to the current best constraints from cosmology. Similarly, for 100 LIGO A+ BNS detections, it is possible to constrain $c_M$ to $\pm0.5$. When analyzing joint $H_0$ and $c_M$ constraints we find that $\sim 400$ LIGO A+ events are needed to constrain $H_0$ to $1\%$ accuracy.
Finally, we discuss the possibility of a nonzero value of $c_M$ biasing standard siren $H_0$ measurements from 100 LIGO A+ detections, and find that $c_M=+1.35$ could bias $H_0$ by 3--4$\sigma$ too low if we incorrectly assume $c_M=0$.

\end{abstract}

\date{\today}
\maketitle


\section{Introduction}\label{sec:introduction}

Einstein's general theory of relativity (GR) is the foundation of gravity. On Solar System scales, not only do its predictions show remarkable agreement with astrophysical data, but precise measurements of phenomena such as the deflection of light around the Sun and the perihelion shift of Mercury rule out many modifications to GR \cite{Will:2014kxa, Berti:2015itd}. Nonetheless, GR exhibits weaknesses at both the very high and the very low energy regimes. At high energies, unavoidable singularities arise during gravitational collapse, and the so-called renormalization problem limits our understanding of quantum gravity \cite{Joshi:2012mk, tHooft:1974toh, Deser:1999mh}. In order to fit observational data on cosmological scales, GR relies on the presence of exotic unknown matter components, namely dark matter and dark energy, to make up $95\%$ of the total energy content of the Universe \cite{Aghanim:2018eyx}. These issues show the current limitations in our understanding of how gravity behaves and interacts with matter in extreme energy regimes. As a result, a number of modified gravity theories have been proposed (see e.g.~\cite{Clifton:2011jh, Koyama:2015vza, Joyce:2014kja} for summaries and reviews), and it is necessary to analyse their consistency, viability, and consequences.

Recently, multimessenger astrophysics has shown great potential for revealing new details of physical phenomena and testing gravity. In particular, combined gravitational wave (GW) and electromagnetic (EM) science is becoming a central topic, allowing us to test different aspects of cosmology and fundamental physics, including alternative dark energy candidates, spatial curvature, the expansion rate of the universe, strong lensing sources, and the graviton mass \cite{Wei:2018cov, Wei:2017emo, Baker:2016reh, Collett:2016dey}, among others. 
For instance, the recent detection of GWs from a binary neutron star (BNS) merger, GW170817 \cite{PhysRevLett.119.161101}, by the advanced Laser Interferometer Gravitational Wave Observatory (LIGO) \cite{LIGO} and advanced Virgo \cite{Virgo}, in conjunction with the detection of an EM counterpart \cite{2041-8205-848-2-L14, 2041-8205-848-2-L15} constrained the propagation speed of GWs, $c_T$, to be $|c_T/c-1|\lesssim 10^{-15}$ relative to the speed of light $c$. This high-precision constraint provides information about possible modifications to GR and strongly disfavors a number of gravity theories proposed in the literature \cite{Baker:2017hug, Creminelli:2017sry, Sakstein:2017xjx, Ezquiaga:2017ekz, Wang:2017rpx, Lombriser:2016yzn}.

The detection of GW170817 and its EM counterpart enabled the first standard siren measurement \cite{Abbott:2017xzu} of the current rate of expansion of the Universe, $H_0$, assuming $\Lambda$CDM cosmology and general-relativistic GW propagation. Besides \cite{Abbott:2017xzu}'s result, independent methods have been used to constrain the Hubble constant, falling into two major categories: large-scale and local observations. From the cosmic microwave background (CMB), the {\it Planck}\/ mission \cite{Planck} found $H_0=67.4\pm 0.5$ km/$\text{s}$/$\text{Mpc}$~\cite{Aghanim:2018eyx}, whereas from local observations of Type-Ia supernovae, the SHoES survey \cite{SHoES} found $H_0=73.48\pm 1.66$ km/$\text{s}$/$\text{Mpc}$ \cite{2018ApJ...855..136R}. These two measurements disagree at the $3.5\sigma$ level. Consequently, obtaining independent GW constraints on $H_0$ with $\sim 1\%$ accuracy could substantially improve understanding of this tension and potentially reveal new physics or sources of systematic error. However, if gravity is described by a theory other than GR, an accurate inference of $H_0$ could be hampered by degeneracies with modified gravity effects. Accordingly, in this paper we constrain $H_0$ with combined GW and EM observational data in the context of alternative theories of gravity with modified GW propagation.

The value of $H_0$ can be obtained if we know the redshift of the source and its luminosity distance. The GW amplitude is inversely related to the luminosity distance, and the scaling constant can be directly inferred from the evolution of the gravitational waveform, which depends on the theory of gravity~\citep{Schutz:1986}. While the source's redshift cannot be inferred directly from GWs, for events with EM counterparts it is possible to obtain the redshift from the EM spectrum. Sources of this kind are known as ``standard sirens'' \cite{0004-637X-629-1-15}, after their EM analogues known as ``standard candles''. In this context, the first BNS detection, GW170817, has already enabled the first standard siren measurement, $H_0=70.0^{+12.0}_{-8.0}$ km $\text{s}^{-1}$ $\text{Mpc}^{-1}$ \cite{Abbott:2017xzu}, and \cite{Chen:2017rfc} suggested that a detection of $\sim 100$ BNS events could yield a measurement with percent-level accuracy (see also \citep{Nissanke:2013,Feeney:2018mkj}).
This target may be attainable in as little as five years, as scheduled improvements bring the LIGO and Virgo detectors to their design sensitivity \cite{Chen:2017rfc}. Within the next decade, the addition of two more detectors, KAGRA \cite{KAGRA} and LIGO-India \cite{IndIGO}, to the LIGO-Virgo network, as well as the planned upgrade of advanced LIGO to LIGO A+ \cite{Aplus}, will further increase sensitivity to BNS mergers, which occur at an astrophysical rate of 110 Gpc$^{-3}$ yr$^{-1}$ or greater \cite{Catalog}.

Inspired by modified gravity models that introduce new degrees of freedom as an alternative to dark energy, we analyze how modified propagation of GWs affects the standard siren estimation of $H_0$. In the context of cosmology, some alternative models have indeed been found to affect $H_0$ measurements and reduce the tension between the cosmological $\Lambda$CDM and local $H_0$ values (see \cite{Ezquiaga:2018btd} for a recent review and discussion). Since the estimated luminosity distance depends on the gravitational theory, we expect it to be different in models where the GW propagation changes due to the presence of additional fields. The impact of modified GW propagation on the measurement of $H_0$ was recently considered in \cite{Belgacem:2018lbp}, which studied a specific model where the amplitude of the GWs is damped due to modified gravity effects in addition to the damping caused by the expansion of the universe. Specifically, this work considered a non-local gravity model and analyzed how constraints on $H_0$ are affected by modifications to gravity and to the equation of state of dark energy. 

In this paper, we stay agnostic regarding the specific underlying gravity theory and study $H_0$ in a model-independent manner, although for simplicity we assume the background universe to evolve according to the $\Lambda$CDM model, while allowing for general modifications for cosmological perturbations and gravitational waves. We compare constraints on $H_0$ with and without considering cosmological data, and quantify the impact of damping effects caused by modified gravity. 
In particular, we identify the extra damping effect with an effective running of the Planck mass, as it is typically found in common modified gravity theories, whose time evolution is given by the fractional dark energy density $\Omega_{DE}$. This time evolution will have one constant arbitrary parameter $c_M$, representing the Planck mass rate today, with $c_M=0$ corresponding to GR with a constant Planck mass. The observable consequences of a running Planck mass have been studied in a number of regimes (see e.g.~\cite{Uzan:2010pm} for a review), and its effect on the propagation of gravitational waves has been considered in \cite{ Pettorino:2014bka, Lombriser:2015sxa,  Nishizawa:2017nef, Arai:2017hxj, Amendola:2017ovw, Linder:2018jil}.
We find that external cosmological data from {\it Planck}, in conjunction with GW170817, constrain $c_M$ to be $-81 < c_M < 28$ at the $95\%$ credible level. These constraints are very weak compared to those from cosmological data alone when assuming a specific modified gravity theory, as $c_M$ also affects CMB and structure formation. For instance, for Horndeski theories, current cosmological constraints give $-0.62< c_M< +1.35$ at $95\%$ confidence level \cite{Noller:2018wyv}.

In addition, we consider populations of events and discuss future forecasts for standard sirens with advanced LIGO and LIGO A+. We find that with A+, 100 BNSs with detected EM counterparts can lead to cosmology-independent constraints on $H_0$ with an accuracy of $\sim 3\%$, and on $c_M$ with $\sigma(c_M)\sim 0.9$. From these results we estimate the need for $\sim 400$ events in order to obtain a $1\%$-accurate constraint on $H_0$ in the presence of a running Planck mass, thereby matching current local and cosmological constraints. Furthermore, we find that $H_0$ and $c_M$ are highly degenerate, which highlights the importance of testing for the parameter $c_M$ to avoid biasing the inferred value of $H_0$ by assuming GR (i.e.~$c_M=0$) in a setting where the true gravitational wave physics is described by $c_M$ of $\mathcal{O}(1)$. In particular, we show that if we have a population of 100 events detected by LIGO A+ with $c_M=1.35$, then the inferred $H_0$, assuming $c_M=0$, will typically be $>3 \sigma$ away from the true value. In this case, the true value of $H_0$ would be outside the posterior $99\%$ credible interval due to the misplaced assumption of $c_M=0$.

The paper is structured as follows. In Section \ref{sec:running} we describe how a time-dependent Planck mass modifies the propagation of GWs, as well as other local and cosmological observables. In Section \ref{sec:sirens} we discuss the use of BNS mergers as standard sirens in the context of a running Planck mass. We describe our inference of $H_0$ in Section \ref{sec:method}, and in Section \ref{sec:H0} we show how estimates of $H_0$ change compared to GR. Finally, in Section \ref{sec:discussion} we summarize our results and give an outlook for future work.
We set the speed of light to unity ($c=1$) throughout.


\section{Running Planck mass}\label{sec:running}
Let us start by considering a perfectly homogeneous and isotropic spatially flat cosmological background. In this case, the metric line element in conformal time $\tau$ takes the form of the Friedmann-Robertson-Walker (FRW) solution:
\begin{equation}\label{eq:backg}
ds^2=a(\tau)^2 \left[-d\tau^2+d\vec{x}^2 \right],
\end{equation}
where $a(\tau)$ is the scale factor determining the expansion of the universe. We next suppose that gravity is described by a modified theory with a time-dependent Planck mass. As previously mentioned, alternative theories typically include new degrees of freedom, which can interact non-trivially with the metric to produce an effective running of the Planck mass. In such a case, there is an ambiguity when defining the stress-energy tensor: the new effects can be interpreted as modifying gravity (i.e.~the Einstein tensor $G_{\mu\nu}$), or as modifying the matter content of the universe (i.e.~the stress-energy tensor $T_{\mu\nu}$). However, without loss of generality, we can always adopt the latter perspective and write the background equations of motion in the standard form
\begin{equation}\label{eq:back}
3\mathcal{H}^2M_P^2=a^2(\rho_m+\rho_{\rm DE}),
\end{equation}
where $\mathcal{H}=a'/a$ is the conformal Hubble rate (with $' \equiv d/d\tau$), $M_P$ is the constant Planck mass, $\rho_m(\tau)$ is the energy density of cold dark matter (CDM), and $\rho_{\rm DE}(\tau)$ is the energy density of dark energy, which is to encapsulate all modified gravity effects. According to this definition, each of the fluid energy-density components is separately conserved:
\begin{equation}
\rho_i'+3\mathcal{H}(\rho_i+P_i)=0 , \qquad i=m, \, {\rm DE} .
\end{equation}
We note that the choice to write the background equation as in eq.~(\ref{eq:back}) is arbitrary, and we could have instead defined it with an effective time-dependent mass $M_*(\tau)$ instead of $M_P$ on the left-hand side and a different $\rho_{\rm DE}$ on the right-hand side. However, in that case, the energy-density components would not have been separately conserved, and thus for simplicity we avoid this choice.

Here we have assumed that there is an additional degree of freedom that couples to the metric in a non-trivial way, which leads to an arbitrary $\rho_{\rm DE}$. In contrast, standard matter components such as baryons and photons have been assumed to be minimally coupled to the metric, as usual, and therefore the way they contribute to the background equations is unchanged. Furthermore, their propagation and evolution are determined by standard geodesics in the given metric background. In particular, for perturbations about FRW, light still propagates at speed $c$.

Next, let us consider small cosmological perturbations around this background universe, and write the total metric as
\begin{equation}
g_{\mu\nu}=\bar{g}_{\mu\nu}+h_{\mu\nu} ,   \quad |h| \ll |\bar{g}|  ,
\end{equation}
where $\bar{g}_{\mu\nu}$ is the background FRW metric and $h_{\mu\nu}$ is a linear perturbation whose transverse and traceless part $h^{TT}_{\mu\nu}$ encodes the GW amplitude. The other components of $h_{\mu\nu}$ determine the amplitude of matter energy-density perturbations, which will be ignored for our purposes.

GR is a single metric theory for a massless spin-2 particle, and hence $h^{TT}_{\mu\nu}$ propagates two physical degrees of freedom corresponding to two tensor polarizations. We consider modified gravity theories that do not propagate additional tensor modes, and therefore $h^{TT}_{\mu\nu}$ alone carries all the information on the evolution of the metric polarizations. Generically, such theories have the following quadratic action
\begin{equation}\label{ActionTensors}
S=\frac{1}{2}\int d^3x d\tau \, M_*^2 a^2\left[ h_A^{'2}  - c_T^2( \vec{\nabla} h_A)^2\right] ,
\end{equation}
where we have expanded $h^{TT}_{\mu\nu}$ into two independent polarization components $h_A$ with $A=+,\times$. The equation of motion for GWs will thus be given by
\begin{equation}\label{eq:gw}
h^{''}_A+\left[ 2+\alpha_M(t)  \right]\mathcal{H}h^{'}_A+k^2c_T^2h_A=0,
\end{equation}
where we have transformed $h_A$ to the spatial 3D Fourier space; its amplitude therefore implicitly depends on time $\tau$ and wavenumber $k$. Here, $\alpha_M$ has been defined as
\begin{equation}
\alpha_M\equiv \frac{d\ln (M_*/M_P)^2}{d\ln a}=\frac{2}{\mathcal{H}}\frac{M_*^{'}}{M_*}.
\end{equation}
We recover GR when $M_*=M_P$ (i.e.~$\alpha_M=0$) and $c_T=1$. However, in modified gravity theories, additional gravitational fields such as scalars or vectors will generically have a time-dependent solution in cosmological backgrounds, which can induce a running of the Planck mass $M_*(\tau)$ due to conformal couplings, even in a local frame. Additional fields can also modify the propagation speed of GWs $c_T(\tau)$ due to non-minimal and non-conformal couplings. We remark that, in this background, both $M_*$ and $c_T$ are functions of time only, and are thus isotropic and polarization-independent. 

Note that we have not added any non-derivative terms to eq.~(\ref{ActionTensors}). This is a consequence of our assumption that the graviton remains massless in the modified theory, and propagates only two polarizations, like in GR. Non-derivative terms like $m^2h_A^2$ can appear in massive gravity theories \cite{deRham:2010kj, deRham:2010ik}, but their GWs are described by five different polarization modes, instead of the two modes $h_A$. More complicated models have also been developed, such as ones in which GWs are directly coupled to additional fields; the perturbations of these fields would appear explicitly in eq.~(\ref{ActionTensors}). This is the case of bigravity theories \cite{Hassan:2011zd}, which propagate one massless and one massive graviton. In these models, GWs can oscillate between the two gravitons (by analogy with neutrinos) and lead to distinctive signals in the waveform \cite{Max:2017flc, Max:2017kdc}. A similar phenomenon is present in multi-vector theories with internal SU(2) symmetry \cite{BeltranJimenez:2018ymu}. Nonetheless, we will not consider such cases in this paper, focusing exclusively on gravity theories that lead to the action of eq.~(\ref{ActionTensors}).

Due to the aforementioned constraints on the speed of GWs from GW170817, we will focus only on running Planck mass models where $c_T=1$. Well-known modified gravity theories that lead to eq.~(\ref{ActionTensors}) with $c_T=1$ are scalar-tensor theories in the Horndeski \cite{Horndeski:1974wa, Deffayet:2011gz} and Beyond Horndeski \cite{Gleyzes:2014dya, Zumalacarregui:2013pma} families that have $c_T=1$ \cite{Baker:2017hug}. Their action takes the form
\begin{equation}\label{ScalarAction}
S_s=\int d^4x\, \sqrt{-g}\left[ G_4(\phi)R + K(X, \phi) - G_3(X,\phi)\Box \phi \right],
\end{equation}
where $\phi$ is an additional scalar gravitational field responsible for dark energy, $X=-\nabla_\mu \phi \nabla^\mu \phi/2$ is the kinetic term of the scalar field, and $G_4$, $K$ and $G_3$ are arbitrary functions. In this case, $\alpha_M=G_4(\phi)$. Specific models belonging to this category are quintessence, $f(R)$ gravity, kinetic gravity braiding and Jordan-Brans-Dicke theory \cite{Baker:2017hug}. Ignoring the matter sector, generalizations of the gravitational action in eq.~(\ref{ScalarAction}) can be obtained by performing a disformal transformation of the metric (see e.g.~\cite{Ezquiaga:2017ekz}) to obtain theories belonging to the Degenerate Higher Order Scalar-Tensor theory (DHOST) family \cite{Langlois:2015cwa, Crisostomi:2016czh, Achour:2016rkg}, which retain the same structure for the GW action. See e.g.~\cite{Kase:2018aps,Langlois:2018dxi,Kobayashi:2019hrl} for reviews on the status of scalar-tensor gravity theories after GW170817.

We also mention that vector-tensor theories belonging to the Generalized Proca \cite{Heisenberg:2014rta} family with $c_T=1$ have an action of the form \cite{Baker:2017hug}
\begin{equation}\label{VectorAction}
S_v=\int d^4x\, \sqrt{-g}\left[R + G_2(X, F, Y) + G_3(X)\nabla_\mu A^\mu \right],
\end{equation}
where $A^\mu$ is an additional vector gravitational field responsible for dark energy; $G_2$ and $G_3$ are arbitrary functions of $X=-A^\mu A_\mu/2$, $F=-F_{\mu\nu}F^{\mu\nu}/4$, or $Y=A^\mu A^\nu F_{\mu}^{\alpha }F_{\nu\alpha}$, where $F_{\mu\nu}=\nabla_\mu A_\nu -\nabla_\nu A_\mu $. Since there are no conformal couplings between the vector field and the metric, it is straightforward to see that these models will also have constant $M_*$, and hence no modification will be seen for tensor polarizations (although the evolution of energy-density matter perturbations will be modified) \cite{Amendola:2017ovw}.

The same happens for Lorentz-breaking vector-tensor theories of the Einstein-Aether family \cite{Jacobson:2000xp, Zlosnik:2006sb}. In this case, the vector field is time-like, and hence defines a preferred frame of reference, and the subclass satisfying $c_T=1$ also has constant $M_*$. We emphasize, however, that theories involving multiple vector fields do allow for non-trivial derivative couplings while still maintaining $c_T=1$ \cite{BeltranJimenez:2018ymu}. However, this case is not encompassed by eq.~(\ref{ActionTensors}), as such models have an additional field explicitly coupled to $h_A$. Extended scalar-vector-tensor theories can also have $c_T=1$ and a non-trivial $M_*(t)$; these are encompassed by eq.~(\ref{ActionTensors}) \cite{Heisenberg:2018mxx}. Generalizations to action (\ref{VectorAction}) including higher derivative interactions have been also studied in \cite{Kreisch:2017uet}, where theories with $c_T=1$ and non-trivial predictions for GW astronomy have been found.

Having discussed examples of dark energy theories that can lead to an effective running Planck mass, we now turn to the observable consequences of such a modification. The time variation of fundamental constants has previously been studied in different regimes. Below, we survey its effect on various observables (see e.g.~\cite{Uzan:2010pm} for a detailed review).

\underline{Cosmology:} The effects of a running Planck mass on cosmology have been considered in \cite{Ade:2014zfo, Pettorino:2014bka, Ade:2015rim, Huang:2015srv} among others. The time-dependence of the Planck mass arises due to the presence of additional fields, which can modify the background evolution of the universe through a time-dependent energy density $\rho_{\rm DE}$, in addition to modifying the evolution of perturbations propagating in the background. Early-time background modifications are well constrained by Big Bang nucleosynthesis (see e.g.~\cite{Copi:2003xd}), and the late-time background modifications are usually constrained using simple parametrizations for the equation of state of dark energy, such as $w_{\rm DE}=w_0+(1-a)w_a$, with $w_0$ and $w_a$ two free constants. We recover the standard $\Lambda$CDM model with $w_0=-1$ and $w_a=0$. The tightest constraints come from combined CMB, supernova, and baryon acoustic oscillation (BAO) measurements yielding,  $w_0=-0.961\pm 0.077$ and $w_a=-0.28^{+0.31}_{-0.27}$ \cite{Aghanim:2018eyx}. Since only small non-vanishing $w_a$ are allowed, from now on we will assume that the background evolution is exactly $\Lambda$CDM. This has the added benefit of disentangling effects coming from the modified background and those coming from the evolution of the perturbations.

Nonetheless, even if the background evolution of the universe is unmodified, the evolution of linear cosmological perturbations can still differ from $\Lambda$CDM.
 It is customary to encode the modifications determining the evolution of the energy-density matter perturbations in two parameters, which affect the evolution of the two Newtonian potentials $\Phi$ and $\Psi$ in the metric
\begin{equation}
ds^2=-a(\tau)^2[(1+2\Phi)d\tau^2+(1-2\Psi)d\vec{x}^2],
\end{equation}
which describes the modifications relative to the FRW solution. The first parameter is an effective Newton's strength $G_{\text{eff}}$ such that the Poisson equation becomes
\begin{equation}
\bar{\Box}\Phi=4\pi  G_{\text{eff}}\rho_m \Delta_m,
\end{equation}
where $\bar{\Box}$ is the d'Alembert operator using the covariant derivatives of the background metric (\ref{eq:backg}), $\Delta_m=\delta\rho_m/\rho_m+3Hv_m$ is the comoving gauge-invariant matter perturbation, $\delta \rho_m$ is the energy-density perturbation and $v_m$ is its velocity potential. The effective Newton's strength depends on time only for sub-horizon perturbations with wavenumber $k>aH$.

The second parameter, $\gamma$, describes an effective anisotropic stress which makes the two metric potentials differ:
\begin{equation}
\gamma = \Psi/\Phi .
\end{equation}
The parameter $\gamma$ depends on time only for sub-horizon scales as well.
Matter and metric perturbations behave in the same way as general relativity when $\gamma=1$ and $G_{\text{eff}}=G_N$, where $G_N$ is Newton's constant. In scalar-tensor theories, it can be seen that the running of the Planck mass causes dark energy to cluster, leading to $G_{\text{eff}}\not=G_N$, and also induces anisotropic stress, making $\gamma\not=1$ (see e.g.~\cite{Bellini:2014fua, Lagos:2017hdr}).

These two parameters are independent of the background evolution, and can differ from GR even when $w_{\rm DE}=-1$. In this unmodified background, we can therefore still obtain different predictions, for instance, for the CMB temperature anisotropies, or large scale galaxy distributions. If we knew exactly the gravitational theory leading to these modifications, we could calculate exactly how $\alpha_M$ affects matter perturbations and use EM cosmological data to constrain the running Planck mass (e.g.~\cite{Bellini:2015xja, Noller:2018wyv} for scalar-tensor theories). However, in this paper we will remain agnostic about the underlying gravity theory, and instead constrain $\alpha_M$ using gravitational wave data. 

\underline{Solar System:} The effects of a running Planck mass at Solar system and laboratory scales have been studied in \cite{Williams:2004qba, PhysRevLett.51.1609, Bonanno:2017dcx}. The relevant modification is a time varying Newton's constant, which affects, for instance, the period and radius of planetary orbits.
However, to remain consistent with the remarkable agreement between GR and observations in this regime, modified gravity theories come equipped with a ``screening mechanism'' that hides all modified gravity effects in dense regions, where one recovers $M_*=M_P$ (see e.g.~\cite{Jain:2010ka, Joyce:2014kja} for reviews). The mechanism can be due to the additional field acquiring a large mass in dense environments, and effectively mediating an undetectable short-range force (chameleon mechanism); due to changing its coupling with matter in this regime, becoming negligible (symmetron mechanism); or having dominant non-linear kinetic terms effectively produce a negligible coupling to matter (Vainshtein mechanism), among other possibilities. These screening mechanisms are expected to act in the Solar System as well as clusters of galaxies.
Currently, observations of the Solar System and laboratory experiments constrain variations of the Planck mass to be of order $10^{-3}$ \cite{Bonanno:2017dcx, Williams:2004qba}.

\underline{Gravitational Waves:} The presence of an additional dark energy field may affect the original emission of the waveform from binary mergers as well as GW propagation. In this paper, we will assume that the additional gravitational degrees of freedom will not modify the emitted waveform, and hence we can use GR to predict the emission of the waveform. Consistency checks must be performed in the future for specific dark energy models to make sure that this is the case, as it has been previously shown that even when stationary black hole solutions may not be affected by the dark energy field, dynamical situations may excite it and leave an imprint \cite{Tattersall:2017erk}. Under our assumptions, we can still use compact binaries with EM counterparts as standard sirens.

Regarding the propagation of gravitational waves, modifications can occur even if the original emitted waveform is the same as in GR. In particular, as shown in eq.~(\ref{eq:gw}) and considered previously in \cite{Pettorino:2014bka, Lombriser:2015sxa, Nishizawa:2017nef,Arai:2017hxj, Amendola:2017ovw, Linder:2018jil}, a running Planck mass can affect GW propagation. In this paper, we will assume that the propagation can be modified inside and outside galaxies, where screening will not be active for GWs and hence the effects of the dark energy field become relevant. However, we will assume that today, in our galaxy, the effective Planck mass is given by $M_P$ (although the present-day value of $\alpha_M$ need not vanish).
Multi-messenger observations allow us to place independent constraints on the running of the Planck mass \cite{Lombriser:2015sxa, Arai:2017hxj, Amendola:2017ovw, Linder:2018jil} as well as constraints on the present Hubble constant $H_0$ (see \cite{Belgacem:2018lbp} for a specific model).



\section{Standard sirens}\label{sec:sirens}
In this section, we show explicitly how a running Planck mass affects standard siren measurements. Since we take the running of the Planck mass to be the result of a new dark energy field, we expect $\alpha_M$ to change over cosmological time scales and only affect the late-time universe. We will consider a specific parametrization for $M_*(t)$ satisfying these requirements, and discuss in detail its consequences. 

Following the approach in \cite{Belgacem:2018lbp,Amendola:2017ovw}, we start by canonically normalizing the field $h_A$ in eq.~(\ref{ActionTensors}), and obtain the following equation of motion for the propagation of the two GW polarizations:
\begin{equation}
\hat{h}_A^{''}+\left(k^2-\frac{a_\text{\rm GW}^{''}}{a_\text{\rm GW}}\right)\hat{h}_A=0 .
\end{equation}
Here, $a_\text{\rm GW}$ is an effective scale factor given by $a_\text{\rm GW}(z)=a(z)(M_*(z)/M_P)$, and $\hat{h}_A=a_\text{\rm GW}h_A$ is the canonically normalized amplitude of GWs. For very small wavelengths, namely for $k^2 \gg  a_{\rm GW}^{''}/a_{\rm GW}$ with $a_{\rm GW}^{''}/a_{\rm GW}=[2\alpha_M^{'}\mathcal{H}+2\mathcal{H}'(1+\alpha_M)+\mathcal{H}^2
(2+\alpha_M)^2]/4$,\footnote{Note that this condition limits how large $\alpha_M$ can be, and how quickly it can evolve in time. Typically, we will consider evolution over cosmological times, and $\alpha_M \ll k^2/\mathcal{H}^2$.} the solution for $\hat{h}_A$ simply is a plane wave with constant amplitude. This means that the original metric perturbation $h_A$ is also a plane wave with a decreasing amplitude due to the factor of $1/a_{\rm GW}$. In GR, the amplitude would simply decay as $1/a$, so we can interpret $a_{\rm GW}$ as the effective scale factor felt by GWs due to the combined effects of the background metric and the background's additional degree of freedom. In this case, the present-day observed amplitude $h_A^o $ is given by
\begin{equation}\label{hObs}
h_A^{o}= \frac{M_*(z)}{M_*(z=0)}h_{A,{\rm GR}}^o \propto \frac{M_*(z)}{M_*(z=0)}\frac{1}{d_L(z)}\equiv \frac{1}{d_{\rm GW}},
\end{equation}
where $z$ is the redshift of the source and $h_{A,{\rm GR}}^o$ is the expression for the observed amplitude in GR; $h_{A,{\rm GR}}^o$ decays as $1/d_L$, with
\begin{equation} \label{dLEM}
d_L=\frac{1+z}{H_0}\int_0^z \frac{d\tilde{z}}{\hat{H}(\tilde{z})}
\end{equation}
the luminosity distance. $\hat{H}(z)=H(z)/H_0$ is the normalized Hubble rate, which is explicitly
\begin{equation}
\hat{H}(z)=\sqrt{\Omega_{m,0}(1+z)^3+\rho_{\rm DE}(z)/\rho_c} .
\end{equation}
We have introduced the fractional energy-density parameter $\Omega=\rho/(3H^2M^2)$, and the critical energy density $\rho_c=3H_0^2M_P^2$. We have also used the fact that $a=(1+z)^{-1}$, with $a=1$ today.

The omitted proportionality factor in eq.~(\ref{hObs}) characterizes the emitted waveform (which is a function of the GW frequency, chirp mass, the effective Newton's constant felt by the compact binary, and the equation of state for neutron stars), and we assume that it is the same as in GR.

We note that eq.~(\ref{hObs}) depends only on the value of $M_*$ at the source and observation points, and not on the intermediate evolution.
This is because we can only measure the cumulative change in the amplitude compared to GR, which depends exclusively on the initial and final values of $M_*$. In keeping with the previous section's discussion, we assume that $M_*(z=0)=M_P$, and that $M_*$ depends only on time. As a consequence, the result depends only on the distance to the source galaxy, not on the properties of the host galaxy itself. Furthermore, we assume that the effective Planck mass evolves in the same way as the cosmological one, and hence it does not depend on whether we are inside a galaxy or not.

In order to make a concrete estimation of the effect of the running Planck mass, we need to assume a specific functional form for $M_*$. Since we are interested in modified-gravity models of dark energy, a common time parametrization for $\alpha_M$ is \cite{Bellini:2015xja, Alonso:2016suf}
\begin{equation}\label{AlphaParam}
\alpha_M(z)=c_M\frac{\Omega_{\rm DE}(z)}{\Omega_{{\rm DE},0}},
\end{equation}
where $c_M$ is a free constant parameter and $\Omega_{\rm DE}$ is the fractional dark energy density. So far the background has been kept arbitrary, but we now assume it to be given by a fiducial $\Lambda$CDM expansion history, and thus from now on we assume that $\rho_{\rm DE}$ is constant, and set $\Omega_{{\rm DE},0}$ to the best-fit Planck value \cite{Aghanim:2018eyx}. Note that the parametrization in eq.~(\ref{AlphaParam}) is not well-suited to $f(R)$ models \cite{Linder:2016wqw}, and hence different time evolutions may also be of interest (see for instance \cite{Kreisch:2017uet}). In any case, simple parametrizations with a few free constants are sufficient for the time being, as observational data does not have the constraining power to test more complicated functions \cite{Gleyzes:2017kpi}.

This parametrization assumes that there are no modified gravity effects at early times, but at late times new degrees of freedom come into play and modify the evolution of the universe through $\Omega_{\rm DE}(z)$. Other parametrizations with early-time modifications can also be considered (see for instance \cite{Alonso:2016suf}).

For the specific case of scalar-tensor theories, it is known how $\alpha_M$ affects the effective Newton constant $G_{\text{eff}}$ and the effective anisotropic stress $\gamma$, and thus cosmological data has been used to constrain $c_M$ to the range $-0.62< c_M< +1.35$ at $95\%$ CL \cite{Noller:2018wyv}, which means $c_M$ is allowed to be of order unity. Thus, in contrast to constraints on the propagation speed of GWs, this kind of modified gravity effect is not yet ruled out.
Forecasts for this same parametrization were studied in \cite{Alonso:2016suf}, where it was found that the 1$\sigma$ uncertainties on the parameter $c_M$ may improve by a factor of 5 when taking into consideration future photometric redshift surveys such as LSST \cite{Abell:2009aa} and SKA \cite{Carilli:2004nx}, as well as the Stage IV CMB experiment \cite{Abazajian:2016yjj}.



Regardless of the underlying modified gravity theory causing $\alpha_M$, from standard sirens we can also place constraints on $c_M$ by measuring the difference between the luminosity distance and the GW distance. As shown in  \cite{Belgacem:2018lbp}, according to eq.~(\ref{hObs}) we have that:
 \begin{equation}
\frac{d_{\rm GW}}{d_L}=\frac{M_P}{M_*}=\exp\left\{ \frac{1}{2}\int^z_0 \frac{dz'}{1+z'}\alpha_M(z')  \right\},
\end{equation}
and for the specific parametrization (\ref{AlphaParam}), we explicitly obtain
\begin{equation}\label{dLratio}
\frac{d_{\rm GW}}{d_L}= \exp\left\{ \frac{c_M}{2\Omega_{DE,0}}\ln \frac{1+z}{(\Omega_{m,0}(1+z)^3+\Omega_{DE,0})^{1/3}}  \right\},
\end{equation}
where we have used the assumption that $\rho_{\rm DE}$ is constant. This ratio describes the cumulative difference in the GW amplitude in GR compared to that in modified gravity, if the wave had the same emitted amplitude. In order to illustrate the effect of $c_M$ on this ratio , we Taylor expand this expression for low redshifts $z\ll 1$ to find
\begin{equation}
\frac{d_{\rm GW}}{d_L} \approx 1+\frac{1}{2} c_M z + \frac{1}{8} c_M \left( c_M - 2 - 6 \Omega_{m,0} \right) z^2 + \mathcal{O}(z^3).
\end{equation}
From here we explicitly see that if $c_M=0$, then $d_{\rm GW}/d_L=1$, and the leading-order correction is proportional to $z$, hence no considerable modifications are expected for low-redshift events. 

When estimating $H_0$ from standard sirens, we first obtain $d_{\rm GW}$ from the waveform and the redshift $z$ from the EM counterpart. Then, we combine eqs.~(\ref{dLratio}) and (\ref{dLEM}) to obtain $H_0$. At low redshifts, we have that
\begin{equation}
H_0= \frac{z}{d_{\rm GW}}+z^2\frac{\left[1-(3/4)\Omega_{m,0}+(1/2)c_M\right]}{d_{\rm GW}}+\mathcal{O}(z^3),
\end{equation}
and thus the estimates of $H_0$ in GR and modified gravity would differ by a factor $z^2c_M/(2d_{\rm GW})$ for a given value of $d_{\rm GW}$. We note that the sign of $c_M$ determines whether the measured $H_0$ in the modified gravity model will be larger or smaller than the GR value. In particular, if $c_M<0$, then $c_M$ contributes with a negative term to $H_0$, yielding a smaller $H_0$, and vice versa. 

\begin{figure}[tb]
	\centering
	\includegraphics[scale=0.35]{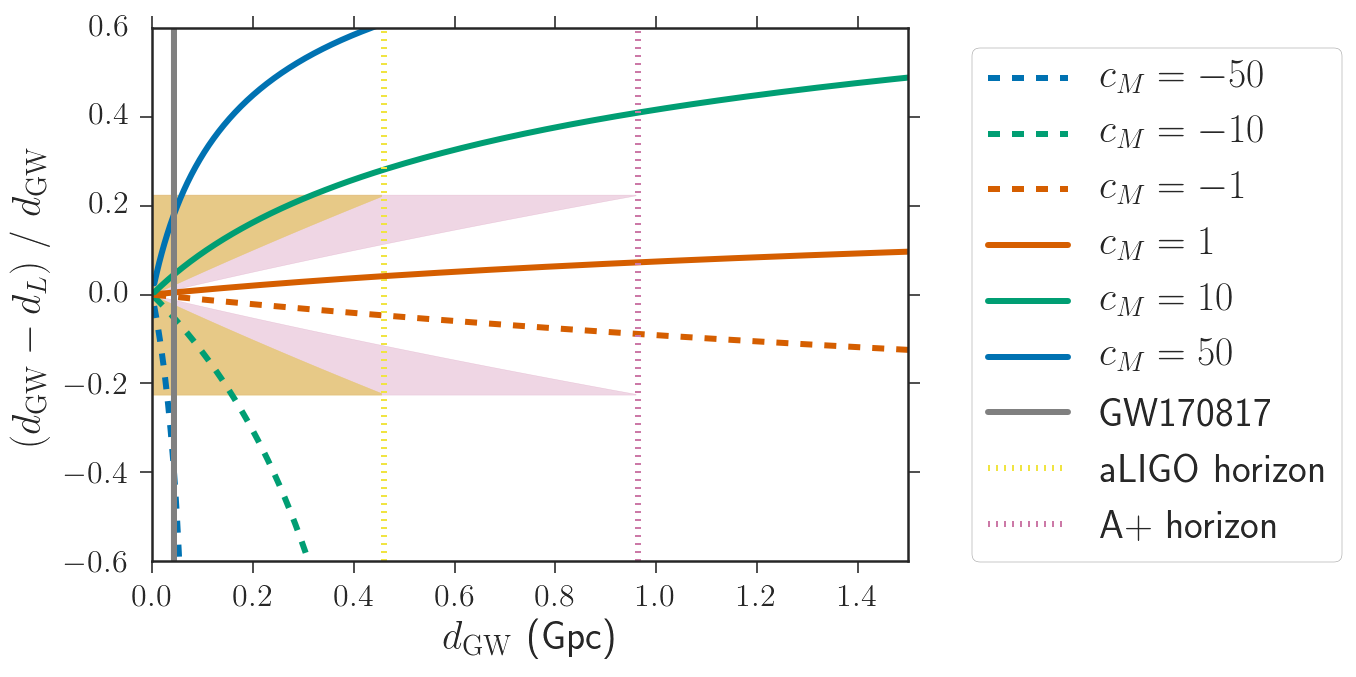}
	\caption{Fractional change in $d_{\rm GW}$ due to modified gravity effects as a function of $d_{\rm GW}$, for different values of $c_M$. Here we have fixed cosmological parameters to the best-fit values from {\it Planck}. The yellow (pink) region correspond to the range of $1\sigma$ $\DGW$ measurement uncertainties for aLIGO (LIGO A+). The sharp cutoff at $\pm 22.5\%$ is due to our assumption that only systems with a single-detector SNR $> 8$ are detected; lower-SNR systems, if included in the sample, will yield broader $\DGW$ measurements (see Sec.~\ref{sec:method}).}
	\label{Fig:dGWvsdLFracError}
\end{figure}

\begin{figure}[b]
 \centering
 \includegraphics[scale=0.5]{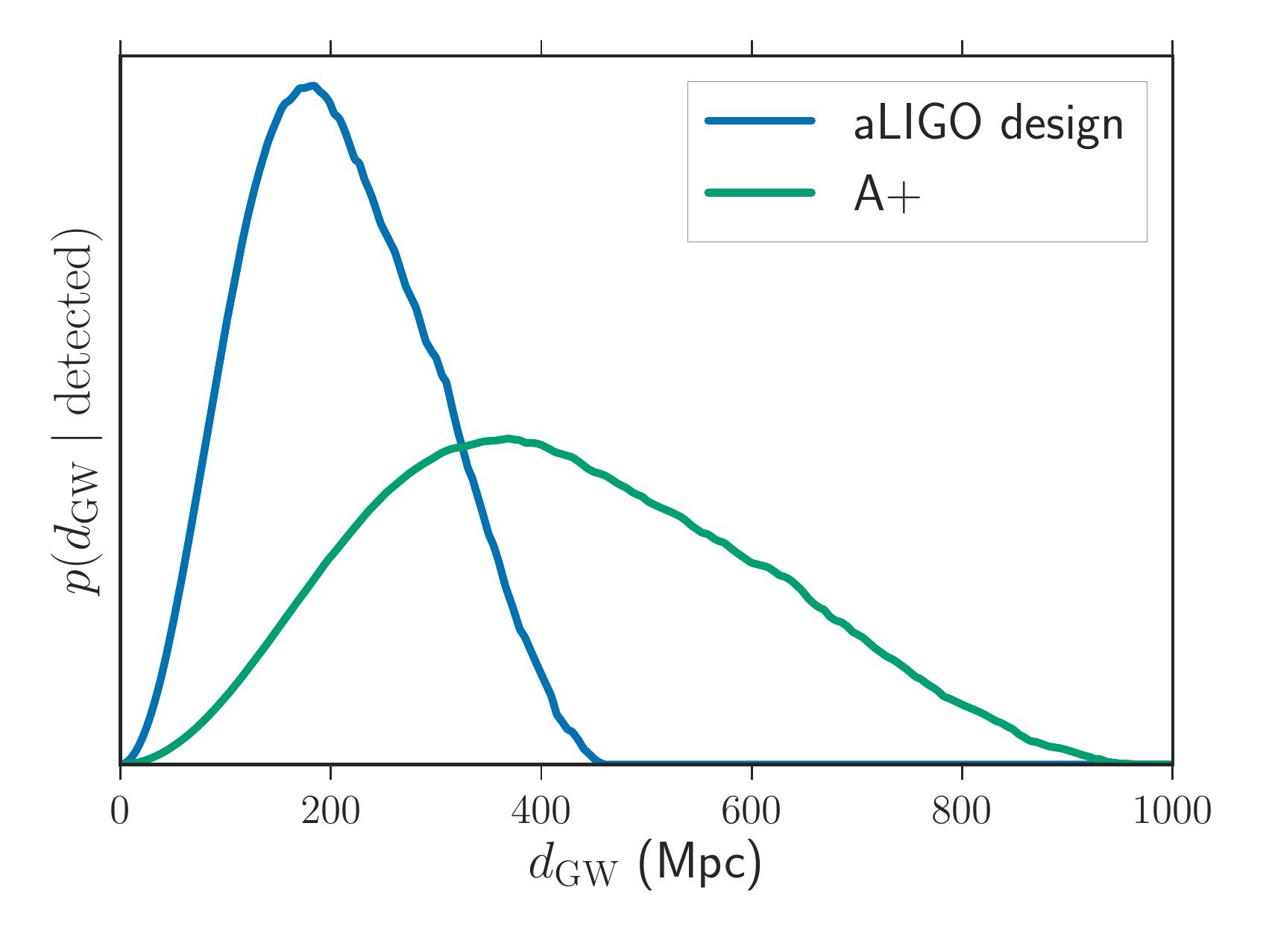}
 \caption{Distance distribution of the BNS events detected by advanced LIGO at design sensitivity and LIGO A+, assuming 1.4--1.4 M$_\odot$ mergers and that the underlying merger rate density follows $\frac{dN}{dV_cdt} \propto (1+z)^{2.7}$ and {\it Planck}\/ cosmology.}
 \label{Fig:BNSdistances}
 \end{figure}

Going beyond the low-redshift limit, Figure~\ref{Fig:dGWvsdLFracError} shows the evolution of the fractional difference $(d_{\rm GW}-d_L)/d_{\rm GW}$ as a function of $d_{\rm GW}$ for different values of $c_M$ (blue, green and orange lines), up to a GW distance of $1.5$ Gpc. For comparison, the GW distances for BNSs detected by aLIGO at design sensitivity and A+ are expected to follow the distributions in Figure~\ref{Fig:BNSdistances} under the assumptions discussed in Section~\ref{sec:method}. The ratio $(d_{\rm GW}-d_L)/d_{\rm GW}$ illustrates the fractional cumulative difference in the GW amplitude that would be detected in GR compared to that detected in modified gravity, for a source at a given distance with the same emitted amplitude. As expected, the larger the $|c_M|$, the larger the fractional difference between the GW distance and $d_L$, indicating a larger deviation from GR. We note that for values of $c_M \gtrsim -3$, the ratio flattens out at large distances due to the fact that our $\alpha_M$ in eq.~(\ref{AlphaParam}) decreases with redshift (or, equivalently, with distance), and hence the accumulated difference between GR and modified gravity becomes negligible at large distances. The distance at which a given curve starts flattening out depends on the value of $c_M$ and the cosmological parameters. Explicitly, from eq.~(\ref{dLratio}), we find that the maximum fractional difference is given by:
\begin{equation}
\lim_{z\rightarrow \infty} \frac{|d_{\rm GW}-d_L|}{d_{\rm GW}}=\left|1-\exp\left\{ -\frac{c_M}{2\Omega_{DE,0}}\ln \Omega_{m,0}^{-1/3}   \right\}  \right|.
\end{equation}
Using best-fit {\it Planck}\/ cosmological parameters, we find that if $c_M=1$, for instance, then the maximum fractional difference will be about $32\%$. However, for sources at aLIGO or LIGO A+ horizons (dotted vertical lines, corresponding to the largest detectable distances in Figure~\ref{Fig:dGWvsdLFracError}), the fractional difference will be about $4\%$ or $7\%$, respectively, for $c_M=1$.


Due to the aforementioned behavior, we conclude that more distant events have \emph{a priori} more constraining power on $c_M$ than nearby events; however, this effect starts to diminish for events at distances much greater than a few Gpc. We note that the measurement uncertainties in $d_{\rm GW}$ tend to grow with distance, as the signal to noise ratio (SNR) of a GW signal scales inversely with the source distance.
The range of expected $1\sigma$ uncertainties in the measured GW distance is shown as solid colored regions in Fig.~\ref{Fig:dGWvsdLFracError}: yellow for aLIGO and pink for LIGO A+, under the simplified assumptions discussed in Section~\ref{sec:method}. The boundary closest to zero of each colored region corresponds to the best sources, which produce the optimal SNR and the smallest distance uncertainties, and thus the tightest constraints on $c_M$.
In the above discussion, we assumed that the uncertainty in the EM distance, $d_{L}$, is negligible, and that all the uncertainty in the fraction $(d_{\rm GW}-d_L)/d_{\rm GW}$ comes from $d_{\rm GW}$. Indeed, for a fixed cosmology, the uncertainty on $d_L$ comes exclusively from the peculiar velocity of the source, which is typically around $150$--$250$ km/s at all distances, and therefore for a fixed background cosmology the fractional error of the luminosity distance $\sigma_{d_L}/d_L$ decreases with distance.

From Figure~\ref{Fig:dGWvsdLFracError}, we see that in order to distinguish GR from a nonzero $c_M$, we need to measure $d_\mathrm{\rm GW}$ to a precision better than the deviation caused by the nonzero value of $c_M$. For a single event, this is only possible for the most extreme values of $c_M$, because we will rarely get an event with a $1\sigma$ distance uncertainty that is comparable to the deviation caused by $-1<c_M<1$. However, by combining a population of events, it will be possible to place tight constraints on $c_M$, as discussed in Section~\ref{sec:population}. A population of BNS events detected by LIGO A+ is especially promising, because the typical source detected by A+ at a given distance has a much higher SNR, and therefore a smaller relative distance uncertainty, than the typical source detected by aLIGO. However, it becomes less useful to extend the detection horizon much beyond A+, because the effect of nonzero $c_M \sim \mathcal{O}(1)$ starts to saturate at $\sim 1.5$ Gpc (additionally, it is increasingly difficult to find counterparts and identify host galaxies for events at high redshift).



Finally, we recall that we have a schematic relationship between $d_{\rm GW}$ and redshift $z$ of the form
\begin{equation}\label{SchemRatio}
d_{\rm GW}=d_L(z,H_0, \Omega_{m,0})R(z, c_M, \Omega_{m,0})
\end{equation}
where $R$ is a function corresponding to the ratio of $d_{\rm GW}/d_L$ given on the righthand side of eq.~(\ref{dLratio}). It is clear that, given cosmological parameters ($H_0$, $\Omega_{m,0}$), we can constrain $c_M$ by measuring $d_{\rm GW}$ and $z$. We emphasize that if the cosmological parameters are fixed by some cosmological data, then the resulting constraint on $c_M$ is not independent of these cosmological data sets. However, if the cosmological parameters $H_0$ and $\Omega_{m,0}$ are taken to be free constants, then eq.~(\ref{SchemRatio}) can be used to find joint constraints on $c_M$, $H_0$ and $\Omega_{m,0}$ which are completely independent from the cosmological data sets. Note that the constraints would still depend on the cosmological model (assumed to be $\Lambda$CDM here), but not on the best-fit values from external data. If $c_M=0$, then $R=1$, and we can use this relationship to find independent constraints on the cosmological parameters, as has been done for $H_0$ in \cite{Abbott:2017xzu}.

While for low-redshift sources $\Omega_{m,0}$ has a negligible effect and the only relevant cosmological parameter is $H_0$, for high-redshift sources $\Omega_{m,0}$ does become relevant and must be taken into consideration. For simplicity, in the rest of the paper we will always fix $\Omega_{m,0}=0.315$, the best-fit value from {\it Planck}\/ 2018 \cite{Aghanim:2018eyx}, while keeping $H_0$ free. In this sense, the constraints that we quote are not fully cosmology-independent, although we have checked that if we free $\Omega_{m,0}$ and use a flat prior with a $6\sigma$ width around the best-fit {\it Planck}\/ value, i.e.~ $0.330<\Omega_{m,0}<0.372$, our results are unaffected, as the uncertainty on $\Omega_{m,0}$ is subdominant, affecting the distance-redshift relation to less than $1\%$ over the redshift range of interest ($z \lesssim 0.1$ for aLIGO and $z \lesssim 0.2$ for LIGO A+) for BNSs.
For this reason, in the rest of the paper we will focus on the joint constraints for $c_M$ and $H_0$ only, and these will be referred to as cosmology-independent constraints. In particular, we will analyze the data from GW170817 as well as forecasts for advanced LIGO and LIGO A+.


\section{Method}\label{sec:method}
In this section, we describe the method used for our standard siren inference of $H_0$ in the context of a running Planck mass. Throughout the analysis, we fit only for $c_M$ and $H_0$, assuming that the other cosmological parameters in the $\Lambda$CDM background (namely, $\Omega_k$, $\Omega_m$ and $\Omega_\Lambda$), collectively denoted by $\Xi$, are known to a few percent. We note that $c_M$ is not expected to correlate with these other cosmological parameters in their impact on CMB observables \citep{Huang:2015srv}, and it is therefore self-consistent to fix the background cosmology to the {\it Planck}\/ 2018 values while measuring $c_M$ with standard sirens. Allowing these parameters to vary by up to 10\% from their best-fit {\it Planck}\/ 2018 values has a $\lesssim 1\%$ effect on the distance-redshift relation over the detectable redshift range, and so even if the current measurement errors were several times larger, marginalizing over the extra uncertainty would have a negligible impact on our results. Within the current {\it Planck}\/ 2018 uncertainties, the distance-redshift relation only varies by $<0.05\%$  over the detectable redshift range. In the future, however, especially if standard sirens with counterparts are detectable to much higher redshifts $z > 1$ by e.g. LISA, one could carry out a joint CMB-standard siren analysis that would incorporate all cosmological parameters.

\underline{GW Measurements}:
We assume that the GW measurement uncertainty of distance scales as $1.8/\rho$ (at $1\sigma$) where $\rho$ is the single-detector SNR of the source. We assume a threshold SNR of $\rho_\mathrm{th} = 8$ for detection. From Fisher matrix arguments, we expect the GW distance to scale inversely with the SNR, with the proportionality factor $>1$ because of the distance-inclination degeneracy~\cite{1994PhRvD..49.2658C}. We choose the $1\sigma$ uncertainty of $1.8/\rho$ to match the expected $H_0$ convergence rate of $(13\%$--$15\%)$/$\sqrt{N}$, where $N$ is the number of GW detections~\cite{Chen:2017rfc}. For simplicity, when simulating a mock population of sources, we assume that the GW distance likelihood is approximated by a Gaussian distribution:
\begin{equation}
p(\DGW^{\rm obs} \mid \DGW) = N_{[\mu = \DGW,\sigma = 1.8/\rho]}(\DGW^{\rm obs})
\end{equation}
where $N_{[\mu, \sigma]}$ denotes the standard normal distribution with mean $\mu$ and standard deviation $\sigma$.
While this is not a realistic approximation for individual sources, when combining tens to hundreds of detections, it yields the expected convergence rate for the recovered cosmological parameters. 

\underline{Likelihood}:
We denote the GW data by $\xGW$ and the EM data by $\xEM$.
The likelihood of the GW data depends on the source's GW distance $\DGW$, sky position $\omega$, inclination $\iota$, and all other parameters of the signal, including its redshift $z$ (which affects the frequency of the waveform), and the source-frame masses, spins, etc., which we collectively denote by $\xi$. The likelihood given the extrinsic parameters (sky localization and inclination) is largely independent of the intrinsic parameters~\cite{2016PhRvD..93b4013S,2017ApJ...840...88C}, and we marginalize over these other parameters to get the GW likelihood given its distance and sky position:
\begin{equation}
    p(\xGW \mid \DGW, \omega) = \int p(\xGW \mid \DGW, \omega, \iota, z,\xi) d \iota d z d\xi .
\end{equation}
Meanwhile, the likelihood of the EM data depends on the host galaxy's sky position and luminosity distance, which is related to its cosmological redshift (the redshift it would have if it were in the Hubble flow, corrected for any peculiar velocities) by the standard $\Lambda$CDM relation.

We put a prior $p(z,\omega)$ on the redshifts and sky positions of the host galaxies. We choose this prior to match a merger rate density that is isotropic and roughly follows the star-formation rate (see Appendix). To avoid a biased measurement of the cosmological parameters, this prior must match the true redshift distribution of the host galaxies; however, we find that any likely deviation from a uniform merger rate density, including a merger rate that traces the star-formation rate, is too small to cause a noticeable bias. Similarly, deviations from an isotropic distribution of sources on the sky (for example, due to large-scale structure) are largely irrelevant to this analysis unless there are significant correlations between the underlying distribution of sources on the sky and the antenna power patterns of the detectors. Moreover, a significant deviation from the assumed merger rate density will be easily detectable with hundreds of host galaxies with well-measured redshifts and sky positions, and can be used to update our prior accordingly.

The relationship between $d_L$ and $\DGW$ is given in eq.~\eqref{dLratio}, where $d_L$ is given by $z$, $H_0$, and the standard $\Lambda$CDM parameters $\Xi$. We denote the function that returns $\DGW$ given $z$, $H_0$, $c_M$ and $\Xi$ by $\hat{d}_\mathrm{\rm GW}$.

The likelihood for the data given $c_M$ and $H_0$ is
\begin{widetext}
\begin{equation}
\label{eq:likelihood}
p(\xGW,\xEM \mid c_M, H_0) = \frac{\int p(\xGW \mid \DGW = \hat{d}_\mathrm{\rm GW}(z,c_M, H_0, \Xi),\omega)p(\xEM \mid z, \omega)p(z, \omega)p(\Xi)dzd\omega d\Xi}{\beta(H_0, c_M)} ,
\end{equation}
\end{widetext}
where $\beta(H_0, c_M)$ ensures that the likelihood integrates to unity over detectable data sets, and accounts for selection effects in the GW detection and measurement process (see Appendix).

We take $p(\Xi)$ to be the posterior on these parameters from {\it Planck}\/ 2018, although, as noted earlier, we can approximate these as being measured exactly and given by a $\delta$-function centered on their mean values. We also assume that the sky position is measured exactly, and that the redshift uncertainty is small (i.e. spectroscopic redshifts, so that the only significant source of uncertainty in the cosmological redshift is in the peculiar velocity correction, typically around 200 km/s~\citep{2017MNRAS.471.4946W}). For a population where the majority of detected sources are at redshifts $z \gtrsim 0.05$ and the redshift uncertainty is subdominant to the $\DGW$ uncertainty, we can therefore approximate the EM likelihood term, $p(\xEM \mid z, \omega)$ by a $\delta$-function centered at the true redshift and sky position:
\begin{equation}
p(\xEM \mid z, \omega) = \delta(z-z_\mathrm{obs})\delta(\omega-\omega_\mathrm{obs}).
\end{equation}
When analyzing GW170817, a very nearby event at redshift $z \sim 0.01$, we include the peculiar velocity uncertainty in the calculations, taking the EM likelihood to be:
\begin{equation}
p(\xEM \mid z, \omega) = N_{[\mu = z,\sigma=\sigma_z]}(z_\mathrm{obs})\delta(\omega-\omega_\mathrm{obs}),
\end{equation}
where $z_\mathrm{obs}$ is the observed, peculiar-velocity corrected redshift, $\sigma_z$ is the uncertainty, and $\omega_\mathrm{obs}$ is the observed sky position. For this analysis, we also choose priors that match the default priors in \cite{Abbott:2017xzu}: $p(\DGW) \propto \DGW^2$ and $p(H_0) \propto 1/H_0$.

\section{Results}\label{sec:H0}
In this section we analyze the one BNS detection so far, GW170817, as well as a simulated population of BNSs detected by aLIGO at design and A+ sensitivities. 
In each case, we present joint constraints on $c_M$ and $H_0$ and highlight the correlations between them. Section \ref{sec:single} discusses constraints from the single event GW170817, detected at $\DGW \sim 40$ Mpc.
Then, in Section \ref{sec:population} we discuss forecast constraints from populations of standard sirens detected with design-sensitivity LIGO and A+.

\subsection{Single Event}\label{sec:single}
\begin{figure}[tb]
	\centering
	\includegraphics[scale=0.5]{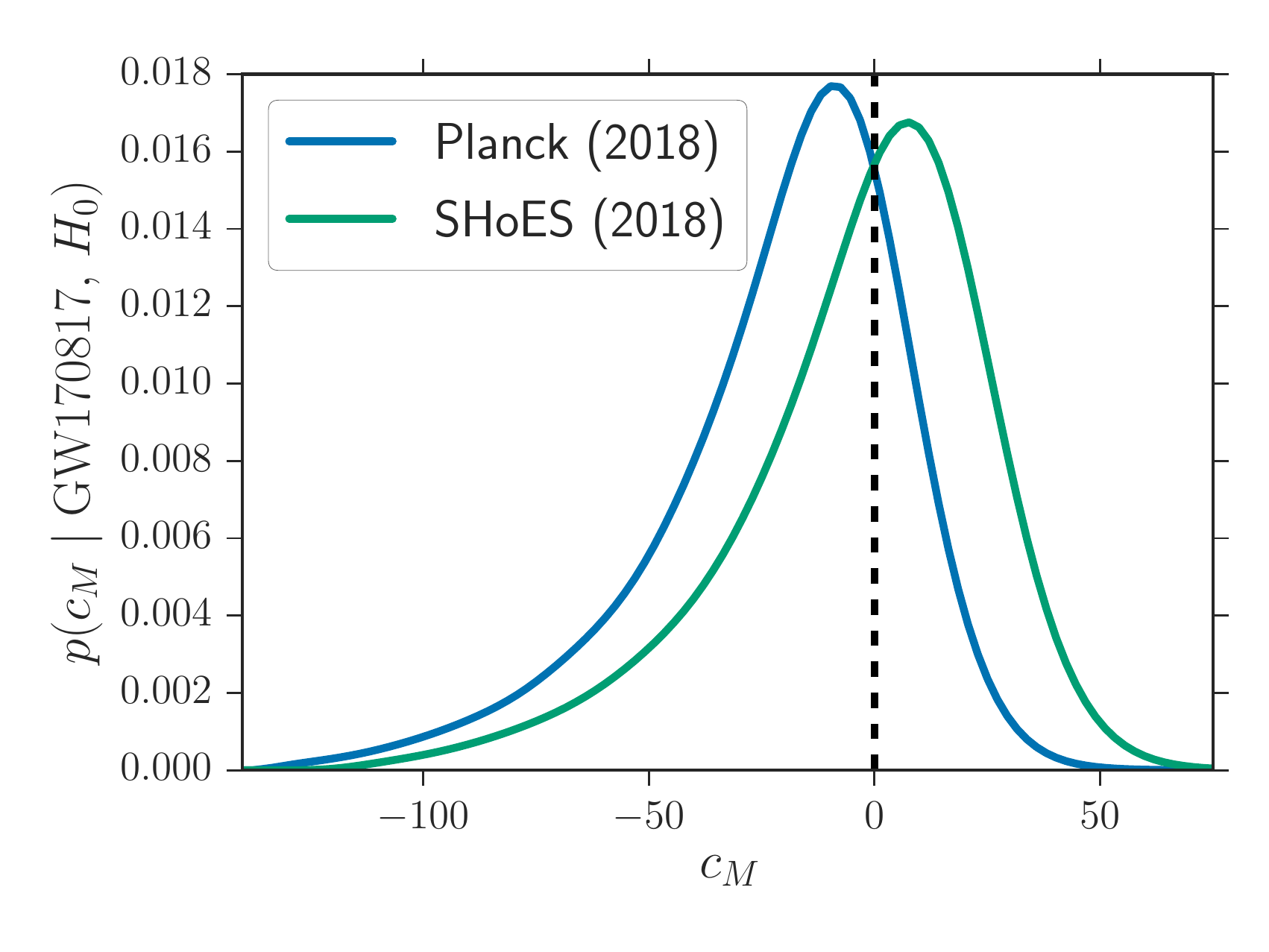}
	\caption{Posterior probability of $c_M$ from the multimessenger detection of GW170817, with an $H_0$ prior given by the {\it Planck}\/ (2018) measurement (blue) and the SHoES (2018) measurement (green).}
	\label{Fig:GW170817_cM}
\end{figure}

In this section we study joint constraints on $c_M$ and $H_0$ from a single multimessenger signal. We start by considering GW170817, which has a measured\footnote{We use the publicly available posterior samples released with~\cite{2018arXiv180511579T} and available at dcc.ligo.org/LIGO-P1800061/public.} GW distance of $\DGW = 41^{+4}_{-7}$ Mpc and a ``Hubble" velocity of $v_H = 3017 \pm 166$ km/s~\cite{Abbott:2017xzu}. Combining the GW distance samples with the redshift of the host galaxy, we can calculate the joint posterior on $H_0$ and $c_M$. If we assume an $H_0$ prior given by {\it Planck}\/ (2018)~\cite{Planck}, we find $c_M = -9^{+21}_{-28}$ (68.3\% credible interval); alternatively, taking a $H_0$ prior given by {\it SHoES}\/ (2018)~\cite{SHoES} gives $c_M = 8^{+21}_{-30}$. The $c_M$ posterior under each assumption is shown in Figure~\ref{Fig:GW170817_cM}. The 95\% credible interval under the {\it Planck}\/ ({\it SHoES}\/) $H_0$ prior is $c_M = -9^{+37}_{-72}$ ($c_M = 8^{+39}_{-74}$). 

We see that even though this event had a high SNR of 32.4, since it was very close-by ($\sim40$ Mpc), the constraints on $c_M$ are very broad. As Figure~\ref{Fig:dGWvsdLFracError} shows, at 40 Mpc, a large range of $c_M$ values would produce GW distances that are consistent with the $\sim 15\%$ distance uncertainty from GW170817. As a comparison, we mention that constraints on $c_M$ have been obtained for scalar-tensor theories from cosmological data, where it was found that $-0.62< c_M< +1.35$ at $95\%$ CL \cite{Noller:2018wyv}. Therefore, current GW constraints allow for $c_M$ of $\mathcal{O}(10)$, whereas cosmological data require $c_M$ of $\mathcal{O}(1)$.

However, future detector networks with improved sensitivity, such as A+, will detect events out to much higher distances with comparable measurement uncertainties. For example, a single event detected by A+ at $\DGW = 400$ Mpc with a $1\sigma$ distance uncertainty of 15\% would constrain $-4 < c_M < 4$ for a true $c_M = 0$, assuming the {\it Planck}\/ $H_0$ measurement as a prior.

\begin{figure}[h]
	\centering
	\includegraphics[scale=0.5]{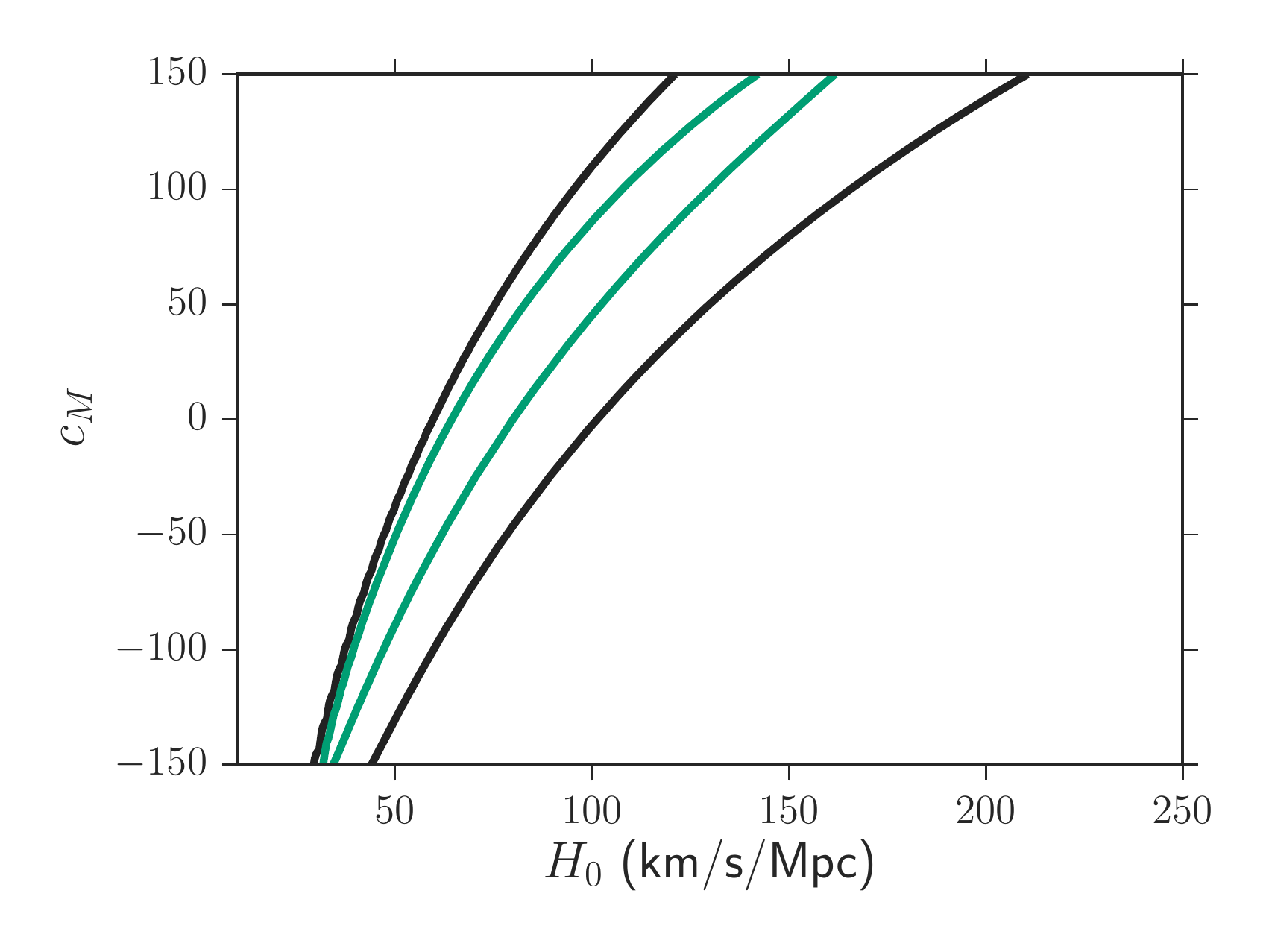}
	\caption{Joint posterior probability of $c_M$ and $H_0$ from GW170817, for a flat prior $-150 < c_M < 150$ and a flat-in-log prior in the range $10 < H_0 < 250$ km/s/Mpc. The black and green contours indicate 90\% and 50\% credibility levels, respectively.}
	\label{Fig:GW170817_cMH0contour}
\end{figure}

We can also obtain constraints on $c_M$ independently of other data sets by using an uninformative prior on $H_0$. In this case, we find a strong positive correlation between $c_M$ and $H_0$, as shown in Fig.~\ref{Fig:GW170817_cMH0contour}.
This correlation arises because a given pair ($z$, $d_{\rm GW}$) can also be achieved in a universe where $H_0$ is larger (and hence the source is closer) but the friction term $c_M$ is correspondingly larger too (and hence the amount of amplitude decay during its travel is larger). Equivalently, the same data could also be fitted by a universe with a smaller $H_0$ and a smaller $c_M$.

\begin{figure}[h]
	\centering
	\includegraphics[scale=0.5]{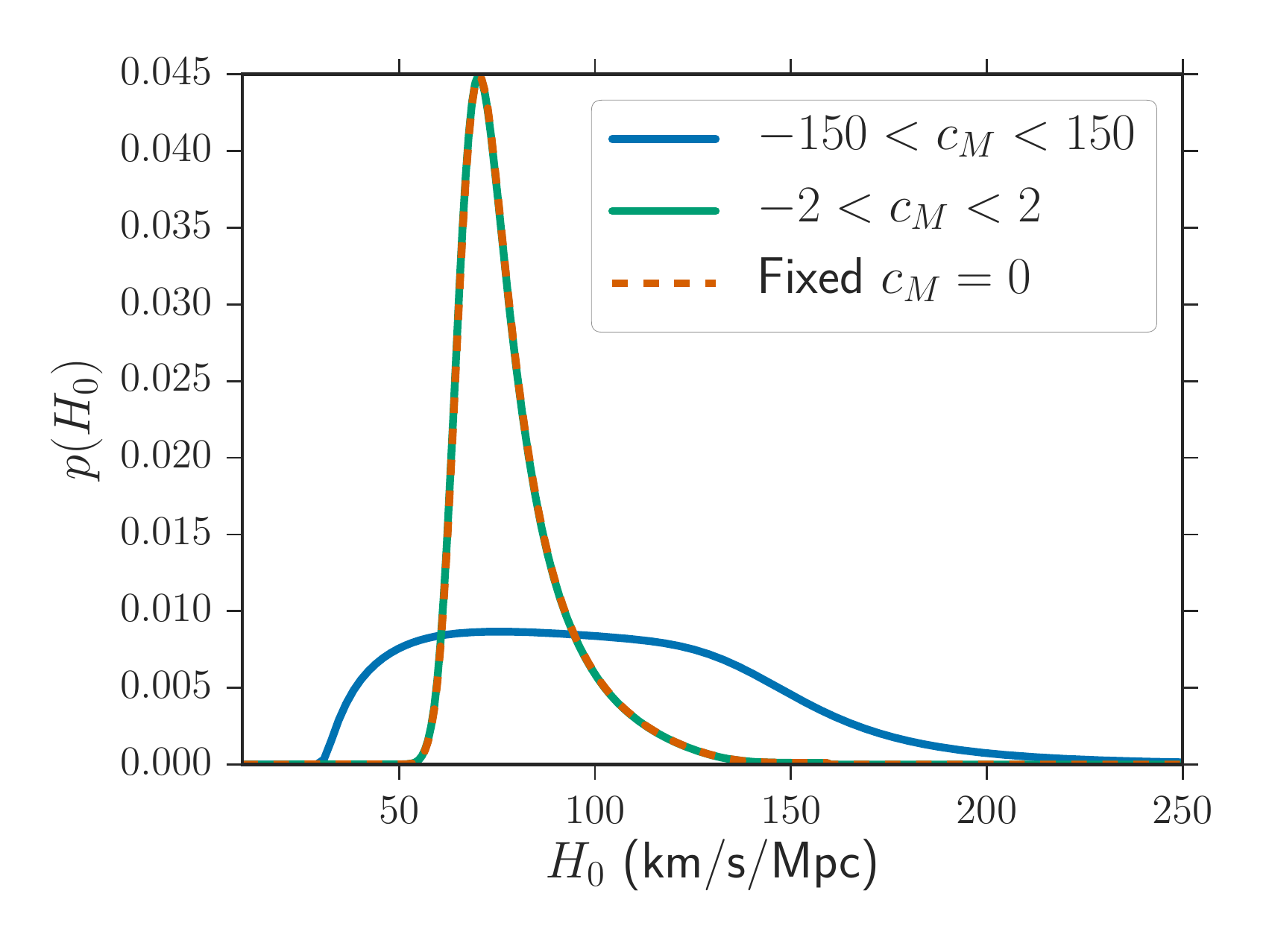}
	\caption{Posterior probability of $H_0$ from GW170817, marginalizing over a wide, flat prior on $c_M \in [-150,150]$ (solid blue line), a narrow prior on $c_M \in [-2,2]$, and constant $c_M=0$ (dashed line; corresponding to GR).}
	\label{Fig:GW170817_margH0_largecMprior_posterior}
\end{figure}

The posterior probability of $H_0$, with $c_M$ marginalized over a flat prior in the range $c_M \in [-150, 150]$ and a flat-in-log prior on $H_0$, is shown in Fig.~\ref{Fig:GW170817_margH0_largecMprior_posterior}. In this case, the constraints become $H_0 = 76^{+53}_{-28}$ km/s/Mpc ($68.3\%$ highest density posterior interval). As a comparison, we also show the constraint on $H_0$ found by marginalizing over a narrow $c_M$ prior, $-2 < c_M < 2$, and assuming $c_M=0$. In both of these cases, we find $H_0=70^{+13}_{-7}$ km/s/Mpc, in agreement with~\cite{2018arXiv180511579T}.
In this case, we find that for this one event the uncertainties on $H_0$ grow by a factor of $>2$ when marginalizing over a very broad prior on $c_M$, but are unaffected by a more reasonable prior $-2 < c_M < 2$.

\subsection{Population}\label{sec:population}
\begin{figure*}
	\centering
	\includegraphics[scale=0.4]{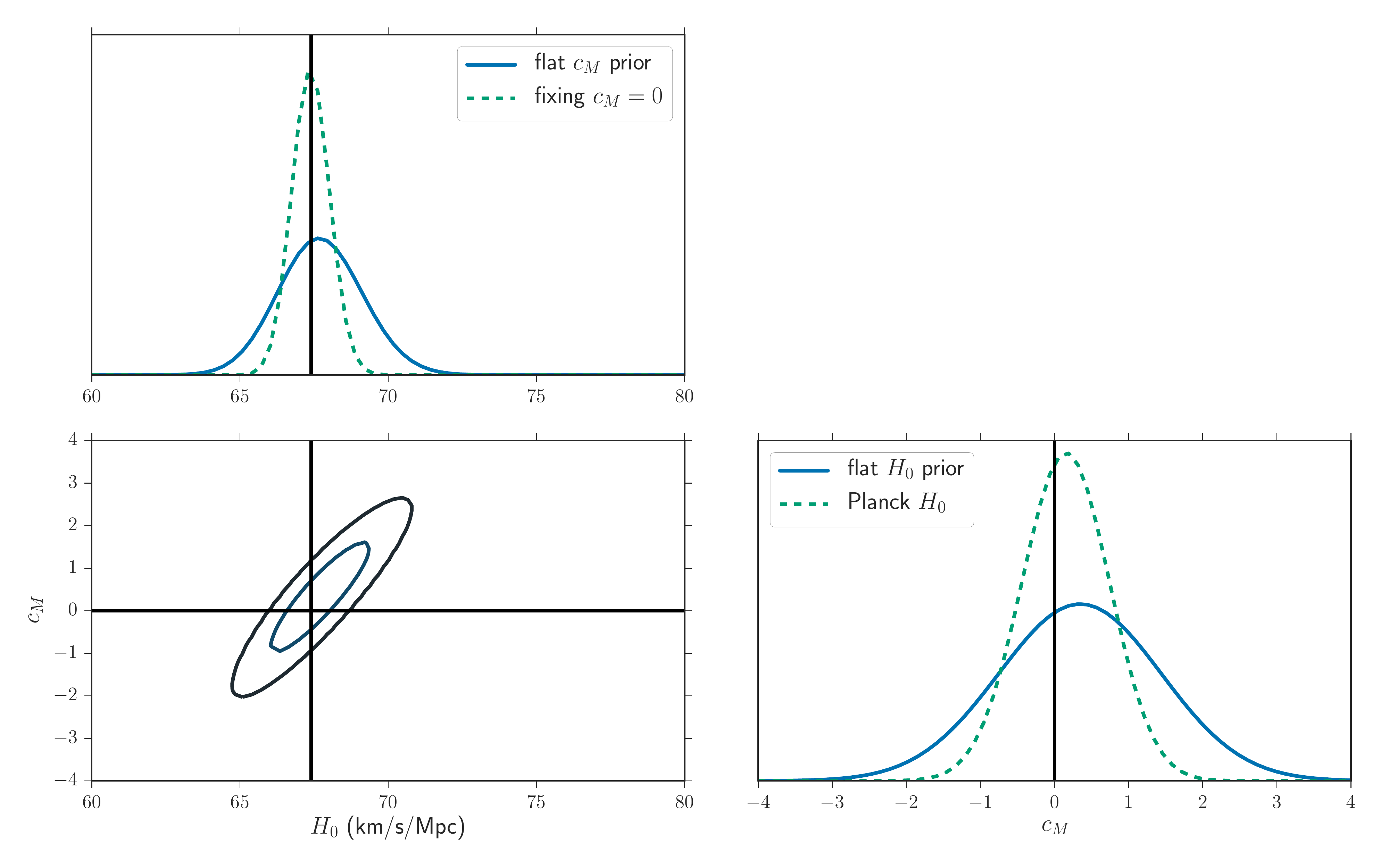}
	\caption{Joint and marginalized posterior probabilities of $c_M$ and $H_0$ (solid blue lines) for 250 mock BNS mergers detected by aLIGO at design sensitivity, with true values of $H_0$ and $c_M$ indicated with solid black lines. The contours in the bottom left panel denote 90\% and 50\% confidence levels. In the top left panel we also show the posterior probability of $H_0$ when fixing $c_M=0$ with a green dotted line. In the bottom right panel we show the posterior probability of $c_M$ when $H_0=67.4\pm 0.5$km/s/Mpc ({\it Planck}\/ 2018). }
	\label{Fig:H0-cM-cornerplot_Nobs250aLIGOcMtrue0}
\end{figure*}

We now consider a population of BNS events detected by aLIGO at design and A+ sensitivities, with EM counterparts that allow us to identify unique redshifts of the host galaxies.
As the GW network sensitivity reaches design sensitivity for aLIGO and later upgrades to A+, a GW event at a given distance will be detected with higher SNR and yield a better-constrained $\DGW$ measurement, meaning that although GW170817 is only sensitive to $c_M \sim \mathcal{O}(50)$, a single event detected by aLIGO at design (A+) sensitivity will typically constrain $c_M$ to a $1\sigma$ width of $\lesssim 10$ ($\lesssim5$).
 In Fig.~\ref{Fig:H0-cM-cornerplot_Nobs250aLIGOcMtrue0} we show the joint and marginalized constraints on $c_M$ and $H_0$ for a simulated population of 250 BNSs detected by aLIGO, where the injected values are $c_M=0$ and $H_0 = 67.4$ km/s/Mpc (indicated by solid black lines). We assume flat priors on both $c_M$ and $H_0$.

The bottom left panel shows the joint posterior probability on $c_M$ and $H_0$, with the contours indicating 90\% and 50\% credibility levels. As expected, there is a positive correlation between $H_0$ and $c_M$, which leads to a broader recovered posterior on $H_0$ when marginalizing over $c_M$, as opposed to fixing $c_M = 0$ (the correct value in this case). This is shown in the top left panel. When fixing $c_M = 0$, the $1\sigma$ constraints on $H_0$ scale roughly as $(13$--$15\%)/\sqrt{N}$ (for aLIGO and A+), giving a $1\sigma$ interval of $0.6$ for 250 events (green dotted line); however, when marginalizing over a completely uninformative prior on $c_M$, the same number of events yields a $1\sigma$ interval that is twice as broad (solid blue line). This implies that with no external knowledge of $c_M$, it would take four times as many events to reach the same precision in the standard siren $H_0$ measurement, or around 200 events to reach $2\%$ as opposed to only $\sim 50$ events if fixing $c_M = 0$~\cite{Chen:2017rfc}. 

Meanwhile, the bottom right panel shows the posterior probability of $c_M$ when marginalizing over $H_0$ with a flat, broad prior (solid blue line), as well as when fixing the $H_0$ prior to the {\it Planck}\/ (2018) posterior (green dotted line). In the former case, taking the {\it Planck}\/ $H_0$ measurement as a prior, this realization gives $c_M=0.16^{+0.58}_{-0.60}$ ($68.3\%$ credible interval), whereas the latter case, which assumes no external measurement of $H_0$, gives $c_M=0.35^{+1.08}_{-1.10}$. Including an external constraint on $H_0$ reduces the uncertainties on $c_M$ by almost a factor of 2. 

Although Fig.~\ref{Fig:H0-cM-cornerplot_Nobs250aLIGOcMtrue0} shows only a single realization of 250 simulated BNSs detected by design-sensitivity aLIGO, we find that the expected constraints on $c_M$ and $H_0$ and the $1/\sqrt{N}$ scalings are typical across many realizations for aLIGO. Generically, for aLIGO, we find constraints on $c_M$ with a $1\sigma$ width that scale roughly as $\sim 9.3/\sqrt{N}$ for an informative $H_0$ prior, and $\sim 16/\sqrt{N}$ for a flat $H_0$ prior. In particular, with 100 BNS events detected by aLIGO, assuming that $H_0$ is obtained from external information (such as from the cosmic microwave background), we would find that $|c_M| \lesssim 0.9$. This number is comparable to the current constraints for scalar-tensor theories obtained from cosmological observations \cite{Noller:2018wyv}.
Similarly, with A+ sensitivity, we find that the same number of events yields constraints on $c_M$ that are tighter by a factor of 2, constraining $c_M$ with a $1\sigma$ width that scales roughly as $\sim 4.7/\sqrt{N}$ for an informative $H_0$ prior, and $\sim 9.5/\sqrt{N}$ for a flat $H_0$ prior. In this case, we find that 100 BNS events would allow us to get a limit $|c_M|\lesssim 0.5$ when assuming $H_0$ is known. 

For the previous example, we chose the true $c_M = 0$ (i.e.~assumed that GWs propagate as predicted by general relativity), and so by fitting a model with an uninformative prior for $c_M$ and $H_0$, we recovered the true values with a larger uncertainty than if we had assumed GR and fixed $c_M = 0$. However, if the true $c_M \neq 0$ but GR is assumed in the usual standard siren analysis, we will recover a biased $H_0$ measurement. Due to the previously shown positive correlation between $H_0$ and $c_M$, if the true $c_M > 0$, the $H_0$ measurement will be biased to low values if falsely assuming $c_M = 0$, and if $c_M < 0$, the $H_0$ measurement will be biased to large values. 
	\begin{figure*}
		\centering
		\includegraphics[scale=0.4]{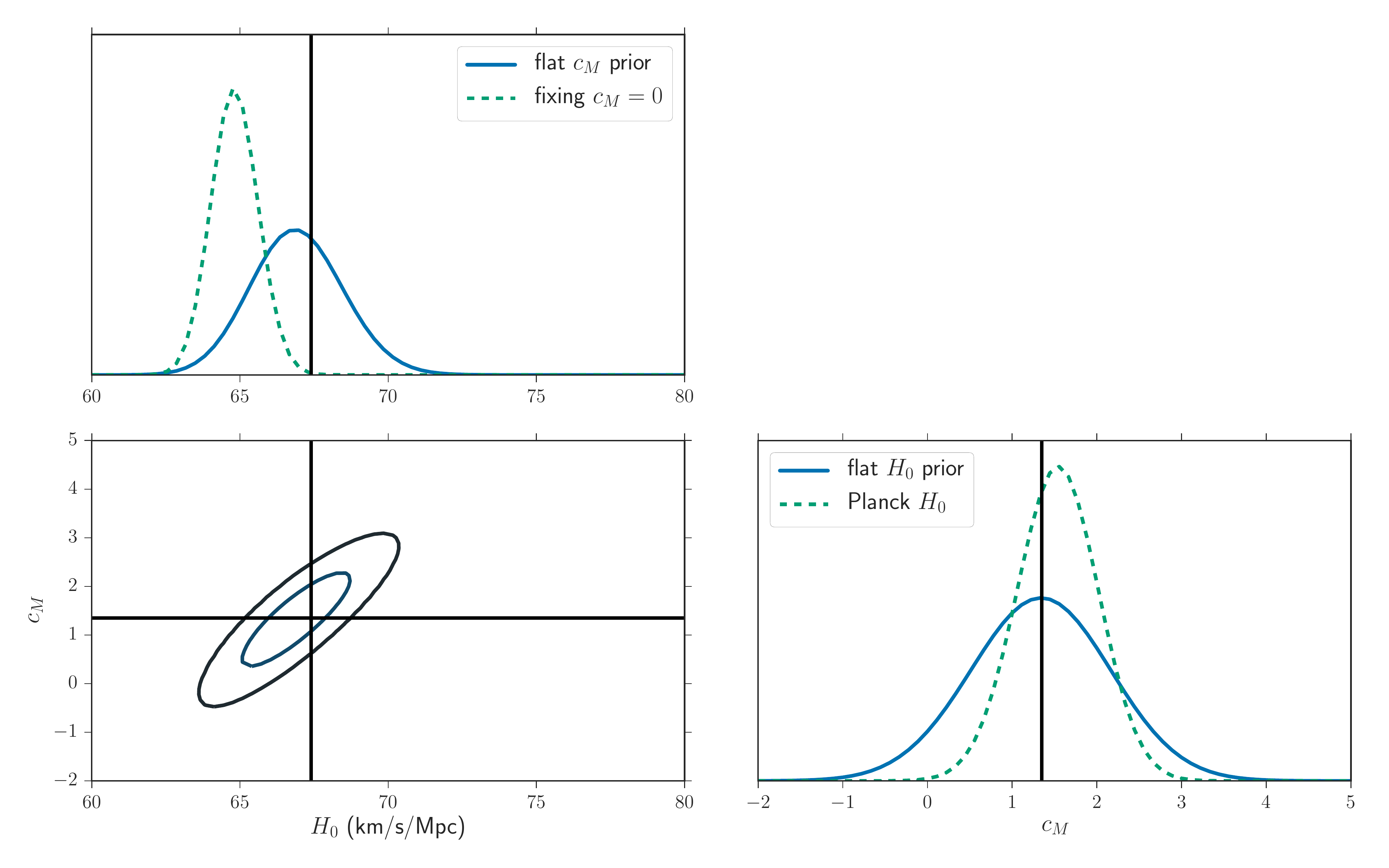}
		\caption{Joint and marginalized posterior probabilities of $c_M$ and $H_0$ (solid blue lines) for 100 mock BNS mergers detected by LIGO A+, with true values of $H_0$ and $c_M$ indicated with solid black lines. The contours in the bottom left panel denote 90\% and 50\% confidence levels. In the top left panel we also show the posterior probability of $H_0$ when $c_M=0$ with a green dotted line. In the bottom right panel we show the posterior probability of $c_M$ when taking the informative prior $H_0=67.4\pm 0.5$km/s/Mpc ({\it Planck}\/ 2018).}
		\label{Fig:H0-cM-cornerplot_Nobs100ApluscMtrue1-35}
	\end{figure*}
As an example, Fig.~\ref{Fig:H0-cM-cornerplot_Nobs100ApluscMtrue1-35} shows the joint and marginalized constraints on $c_M$ and $H_0$ for a simulation in which we injected $c_M = 1.35$. 
We emphasize that this value of $c_M$ is currently allowed within the 95\% confidence interval inferred by~\cite{Noller:2018wyv} for scalar-tensor theories of gravity. For this simulation, we take 100 mock BNS events detected by LIGO A+.

As in the previous figure, in the bottom left panel we show the joint posterior probability, where the contours indicate 50\% and 90\% levels. 
The bottom right panel shows the posterior probability of $c_M$ marginalizing over an uninformative prior on $H_0$ (solid blue line), as well as when adopting the {\it Planck}\/ (2018) measurement as the prior on $H_0$ (green dotted line). In the former case, it is found that $c_M=1.3\pm 0.8$ ($68.3\%$ credible interval), whereas in the latter case  $c_M=1.5 \pm 0.5$.

In the top left panel, we see that the posterior probability of $H_0$ is significantly biased away from its true value if we falsely assume $c_M=0$ (green dotted line). If we properly marginalize over a flat $c_M$ prior, we find $H_0=66.9^{+1.6}_{-1.5}$ km/s/Mpc, whereas fixing $c_M = 0$ gives $H_0=64.8^{+0.8}_{-0.8}$ km/s/Mpc. In the latter case, the true $H_0$ is outside the $99\%$ credible interval. In other words, incorrectly assuming $c_M = 0$ yields an $H_0$ measurement that is biased by more than $3\sigma$ with only 100 events detected by A+. In general, if the true value is positive ($c_M>0$) and we fit a model with $c_M=0$, we bias $H_0$ towards lower values than the true one. Conversely, if the true value is negative ($c_M<0$), we bias $H_0$ towards higher values.


\section{Discussion}\label{sec:discussion}

In this paper we describe a modification to GR which impacts the propagation of GWs. This extension corresponds to a possible running of the Planck mass, which we describe with one free parameter $c_M$, where $c_M=0$ represents GR with a constant Planck mass. This modification affects the friction of GW amplitudes when they propagate through a homogeneous and isotropic universe, and also affects the evolution of matter perturbations in different ways depending on the theory of interest. Here we focus on GW standard siren measurements, studying how a $c_M\ne0$ modification is degenerate with the value of the local Hubble expansion rate, $H_0$. We also explore the ability to constrain both these quantities with future standard siren events detected by LIGO at design and A+ sensitivities.

Studying the event GW170817, we find that if we include external cosmological data, namely the {\it Planck}\/ 2018 $H_0$ posterior, then we find $-81< c_M < 28$ at $95\%$ credibility. This constraint is very weak compared to existing constraints from cosmological data when a specific modified gravity theory is considered. Since $c_M$ affects matter perturbations, it leaves potentially detectable imprints on CMB and structure formation. For Horndeski theories, current cosmological constraints give $-0.62< c_M< +1.35$ at $95\%$ CL \cite{Noller:2018wyv}. Hence, current GW constraints allow for $c_M$ of $\mathcal{O}(10)$, whereas cosmological data require $c_M$ of $\mathcal{O}(1)$. On the other hand, for cosmology-independent results, we find that $c_M$ and $H_0$ are highly degenerate, and constraints on $H_0$ are degraded from $H_0=70^{+12}_{-8}$ km/s/Mpc (when $c_M=0$, that is, when GR is assumed to be correct) to $H_0=76^{+53}_{-28}$ km/s/Mpc (when marginalizing over a very broad $c_M$ prior, $-150 < c_M < 150$).


In addition, we consider populations of events and discuss future forecasts for standard sirens with aLIGO and LIGO A+. We find that 100 BNSs detected by A+ with identified EM counterparts can lead to cosmology-independent constraints on $H_0$ with an accuracy of $\sim 3\%$, and on $c_M$ with $\sigma(c_M)\sim 0.9$ (or $\sigma(c_M) \sim 1.6$ with 100 aLIGO detections). From these results we estimate the need for $400$ detections in order to obtain a $\sim1\%$ constraint on $H_0$, or four times what is required if $c_M$ is known exactly to be zero. Furthermore, we find that $H_0$ and $c_M$ are highly degenerate, which highlights the importance of testing for the parameter $c_M$ to avoid biasing the inferred value of $H_0$ by wrongly assuming GR. In particular, we show that if we have a population of 100 events with $c_M=1.35$, then the inferred $H_0$, assuming $c_M=0$, will be $>3\sigma$ below the true value. In this case, the actual $H_0$ of the population may be ruled out at more than $99\%$ confidence due to the incorrect assumption that $c_M=0$. This result emphasizes the importance of testing the minimal assumptions of one's models. Finding a bias of this magnitude could help arbitrate the current discrepancy between local and cosmological $H_0$ constraints.

It is important to discuss some caveats of our results and calculations. In general, all population results depend on the time-evolution of the background, assumed to be $\Lambda$CDM here. They also depend on the specific values of the cosmological parameters considered, and in some cases we also assumed $H_0$ to be fixed. One possible extension to the analysis made in this paper could involve a change in the background, for example changing $\Lambda$ ($w=-1$) to a more general form of dark energy, $w_{\rm DE}= w_0+w_a(1-a)$.

In addition, the numbers quoted here also depend on the parametrization adopted for $\alpha_M(t)$, which means that even if these constraints are found to be in tension with future measurements, we cannot conclude that all models with nonzero $\alpha_T$ are disfavored, but rather that the specific time behaviour assumed here, $\alpha_M(t)=c_M\frac{\Omega_{\rm DE}(z)}{\Omega_{{\rm DE},0}}$, is disfavored. This is why alternative parametrizations, such as the ones considered in \cite{Noller:2018wyv}, must also be tested.

Furthermore, we have made the crucial assumption that the GW emission of the BNS source is the same in GR as in the modified gravity theory. One typically justifies this by arguing that some compact object solutions, such as those for black holes, are exactly the same as the solutions in GR. However, as shown in \cite{Tattersall:2017erk}, even if the stationary solutions in the strong-field regime of modified gravity look the same as in GR, the emission of GWs in dynamical environments can still differ.
Another argument is that the extra gravitational field may be suppressed due to a screening mechanism in the intermediate and high-energy regimes. However, even if this is true, one should check that the suppressed effects are negligible given LIGO's current measurement uncertainty.
To date, GR waveforms have been found to describe all detected compact binary GWs. Nonetheless, it would be interesting to have analytical or numerical calculations (for instance for Horndeski theories) that allow us to estimate the size of modified gravity effects in the waveform, and determine whether they can be seen with LIGO A+ or next-generation detectors.

Finally, we highlight that given the current constraints on the propagation speed of gravitational waves $|c_T/c-1|\lesssim 10^{-15}$, a constraint on $\alpha_M$ would have a large impact on modified gravity theories, as this is the only possible effect that could be detected with GW data for second-order-derivative scalar-tensor theories, and the only remaining non-trivial extension to GR that can be achieved in this case. For the model in eq.~(\ref{ScalarAction}), a constraint pointing to $\alpha_M\approx 0$ (i.e.~no modification to GW propagation) would disallow all scalar-tensor interactions but those given by
\begin{equation}
S_s=\int d^4x\, \sqrt{-g}\left[ R + K(X, \phi) - G_3(X,\phi)\Box \phi \right].
\end{equation}
In the sub-horizon regime, these models always give $\gamma=1$, as well as $G_{\text{eff}}=G_N$ for structure formation when the coupling term $G_3$ is sufficiently small, and hence modified gravity would only affect scales near the cosmological horizon.



\section*{Acknowledgments}
\vspace{-0.2in}
We would like to thank Ignacy Sawicki for useful discussions during the development of this work. ML, MF, PL, and DEH are supported by the Kavli Institute for Cosmological Physics at the University of Chicago through an endowment from the Kavli Foundation and its founder Fred Kavli. MF was also supported by the NSF Graduate Research Fellowship Program under grant DGE-1746045. PL was supported in part by the Natural Sciences and Engineering Research Council of Canada, and by NSF grant PHY-1505124. MF, PL, and DEH were supported by NSF grant PHY-1708081.

\appendix*
\section{Statistical Formalism}
In this appendix, we give additional details regarding the statistical formalism summarized in Sec.~\ref{sec:method}. In order to arrive at the likelihood (eq.~\ref{eq:likelihood}) for a given EM and GW dataset given values of $c_M$ and $H_0$, we follow the formalism of~\citep{Abbott:2017xzu,Chen:2017rfc,Pardo:2018,Mandel:2018} for incorporating measurement uncertainty and selection effects.
 
In the likelihood eq.~\ref{eq:likelihood}, we include a term $\beta(H_0, c_M)$ to account for selection effects in the measurement process. This term is given by the integral of the numerator over all detectable EM and GW datasets~\citep{Mandel:2018}:
\begin{widetext}
\begin{equation}
\label{eq:beta}
\beta(c_M, H_0) = \int_{\xGW > \xGW^\mathrm{thresh}, \xEM > \xEM^\mathrm{thresh}} p(\xGW \mid \DGW = \hat{d}_\mathrm{\rm GW}(z,c_M, H_0, \Xi),\omega)p(\xEM \mid z, \omega)p(z, \omega)p(\Xi)dzd\omega d\Xi d\xGW d\xEM,
\end{equation}
\end{widetext}
where we assume that a given EM or GW dataset is detected if and only if it is above certain threshold. In reality, the detectability of an EM counterpart to a GW event may depend on details such as the inclination, masses, and apparent magnitude of the source, but for simplicity we assume that all BNS mergers detected in GWs will have an observed EM counterpart and an identified host galaxy. If such counterparts are similar to the kilonova associated with GW170817, their detection is certainly feasible with current telescopes for aLIGO sources (which will be at redshifts $z \lesssim 0.1$), and with future telescopes such as LSST for higher-redshift A+ sources~\cite{Chen:2017rfc}. We therefore assume that the integral over $\xEM$ is independent of the other terms, and ignore it. However, if it becomes the case with future detections that only a subset of GW BNS events have identified host galaxies,this term must be modeled and incorporated into the likelihood.

For the GW selection effects, we assume that a BNS is detected if it produces a single-detector SNR $\rho > 8$. When a real population of BNSs is detected in GWs, this assumption can be easily modified to consider the network SNR, or calibrated to injection campaigns in real data~\citep{O2pop}. As in~\citep{Chen:2017rfc}, we define:
\begin{equation}
P_\mathrm{det}(\DGW) \equiv \int_{\xGW > \xGW^\mathrm{thresh}}p(\xGW \mid \DGW,\omega)p(\omega)d\omega d\xGW.
\end{equation}
(Recall that our assumed prior $p(z, \omega)$ is separable, $p(z, \omega) = p(z)p(\omega)$.) We evaluate the term $P_\mathrm{det}(\DGW)$ with the procedure described in~\citep{Chen:2017rfc}. In particular, we make the simplifying assumptions that the detectability of a GW waveform is independent of its redshift, which affects the observed frequency and therefore the SNR of the source, but only by a negligible amount for the redshifts $z \lesssim 0.2$ considered here. We assume that all BNS sources are nonspinning (the dimensionless spin is expected to be very small, $a < 0.05$, for BNS sources) and 1.4--1.4 $M_\odot$ in mass. The mass distribution will, in general, affect the term $P_\mathrm{det}(\DGW)$, since the SNR of a GW source is a strong function of the binary's mass. However, to leading order the SNR depends only on the chirp mass, and so the term $P_\mathrm{det}(\DGW)$ depends only on the underlying distribution of BNS chirp masses. The chirp mass of each source is measured extremely well, and so with $\mathcal{O}(100)$ sources, the distribution of chirp masses will be accurately determined and can be used to update the function $P_\mathrm{det}(\DGW)$ used in standard siren analyses.  Note that we also assume that any running of the Planck mass (nonzero $c_M$) only affects the amplitude of the signal and not the frequency evolution of the waveform; otherwise, the recovered masses would be affected. 

We therefore have that the $\beta(c_M,H_0)$ term in the likelihood is given by:
\begin{widetext}
\begin{equation}
\beta(c_M, H_0) = \int P_\mathrm{det} (\DGW = \hat{d}_\mathrm{\rm GW}(z,c_M, H_0, \Xi))p(z)p(\Xi)dzd\omega d\Xi.
\end{equation}
\end{widetext}
The prior on the redshift of the source, $p(z)$, enters into this equation. In general, to avoid a biased measurement, the prior $p(z)$ must match the true redshift distribution. In our simulations, we assume that the underlying redshift distribution matches a merger rate that roughly traces the low-redshift star-formation rate:
\begin{equation}
p(z) = \frac{dV_c}{dz}\frac{1}{1+z}\left( 1+z \right)^{2.7},
\end{equation}
where $V_c$ is the comoving volume, $(1+z)^{2.7}$ approximates the Madau-Dickinson star-formation rate~\citep{2014ARA&A..52..415M} at low redshift and the factor of $\frac{1}{1+z}$ accounts for difference in clocks between the source-frame and the detector-frame. In reality, for the redshifts considered here, $z \lesssim 0.2$, any reasonable redshift distribution, including one that traces the star-formation rate, is a very small deviation from the uniform-in-comoving volume and source-frame time redshift distribution, and is unlikely to significantly affect the results. Furthermore, the true redshift distribution will be accurately measured given a precise redshift measurement of each identified host galaxy, and can be used to update the prior $p(z)$.

\bibliography{RefModifiedGravity}

\begin{thebibliography}{95}%
\makeatletter
\providecommand \@ifxundefined [1]{%
 \@ifx{#1\undefined}
}%
\providecommand \@ifnum [1]{%
 \ifnum #1\expandafter \@firstoftwo
 \else \expandafter \@secondoftwo
 \fi
}%
\providecommand \@ifx [1]{%
 \ifx #1\expandafter \@firstoftwo
 \else \expandafter \@secondoftwo
 \fi
}%
\providecommand \natexlab [1]{#1}%
\providecommand \enquote  [1]{``#1''}%
\providecommand \bibnamefont  [1]{#1}%
\providecommand \bibfnamefont [1]{#1}%
\providecommand \citenamefont [1]{#1}%
\providecommand \href@noop [0]{\@secondoftwo}%
\providecommand \href [0]{\begingroup \@sanitize@url \@href}%
\providecommand \@href[1]{\@@startlink{#1}\@@href}%
\providecommand \@@href[1]{\endgroup#1\@@endlink}%
\providecommand \@sanitize@url [0]{\catcode `\\12\catcode `\$12\catcode
  `\&12\catcode `\#12\catcode `\^12\catcode `\_12\catcode `\%12\relax}%
\providecommand \@@startlink[1]{}%
\providecommand \@@endlink[0]{}%
\providecommand \url  [0]{\begingroup\@sanitize@url \@url }%
\providecommand \@url [1]{\endgroup\@href {#1}{\urlprefix }}%
\providecommand \urlprefix  [0]{URL }%
\providecommand \Eprint [0]{\href }%
\providecommand \doibase [0]{http://dx.doi.org/}%
\providecommand \selectlanguage [0]{\@gobble}%
\providecommand \bibinfo  [0]{\@secondoftwo}%
\providecommand \bibfield  [0]{\@secondoftwo}%
\providecommand \translation [1]{[#1]}%
\providecommand \BibitemOpen [0]{}%
\providecommand \bibitemStop [0]{}%
\providecommand \bibitemNoStop [0]{.\EOS\space}%
\providecommand \EOS [0]{\spacefactor3000\relax}%
\providecommand \BibitemShut  [1]{\csname bibitem#1\endcsname}%
\let\auto@bib@innerbib\@empty
\bibitem [{\citenamefont {Will}(2014)}]{Will:2014kxa}%
  \BibitemOpen
  \bibfield  {author} {\bibinfo {author} {\bibfnamefont {C.~M.}\ \bibnamefont
  {Will}},\ }\href {\doibase 10.12942/lrr-2014-4} {\bibfield  {journal}
  {\bibinfo  {journal} {Living Rev. Rel.}\ }\textbf {\bibinfo {volume} {17}},\
  \bibinfo {pages} {4} (\bibinfo {year} {2014})},\ \Eprint
  {http://arxiv.org/abs/1403.7377} {arXiv:1403.7377 [gr-qc]} \BibitemShut
  {NoStop}%
\bibitem [{\citenamefont {Berti}\ \emph {et~al.}(2015)\citenamefont {Berti}
  \emph {et~al.}}]{Berti:2015itd}%
  \BibitemOpen
  \bibfield  {author} {\bibinfo {author} {\bibfnamefont {E.}~\bibnamefont
  {Berti}} \emph {et~al.},\ }\href {\doibase 10.1088/0264-9381/32/24/243001}
  {\bibfield  {journal} {\bibinfo  {journal} {Class. Quant. Grav.}\ }\textbf
  {\bibinfo {volume} {32}},\ \bibinfo {pages} {243001} (\bibinfo {year}
  {2015})},\ \Eprint {http://arxiv.org/abs/1501.07274} {arXiv:1501.07274
  [gr-qc]} \BibitemShut {NoStop}%
\bibitem [{\citenamefont {Joshi}\ and\ \citenamefont
  {Malafarina}(2011)}]{Joshi:2012mk}%
  \BibitemOpen
  \bibfield  {author} {\bibinfo {author} {\bibfnamefont {P.~S.}\ \bibnamefont
  {Joshi}}\ and\ \bibinfo {author} {\bibfnamefont {D.}~\bibnamefont
  {Malafarina}},\ }\href {\doibase 10.1142/S0218271811020792} {\bibfield
  {journal} {\bibinfo  {journal} {Int. J. Mod. Phys.}\ }\textbf {\bibinfo
  {volume} {D20}},\ \bibinfo {pages} {2641} (\bibinfo {year} {2011})},\ \Eprint
  {http://arxiv.org/abs/1201.3660} {arXiv:1201.3660 [gr-qc]} \BibitemShut
  {NoStop}%
\bibitem [{\citenamefont {'t~Hooft}\ and\ \citenamefont
  {Veltman}(1974)}]{tHooft:1974toh}%
  \BibitemOpen
  \bibfield  {author} {\bibinfo {author} {\bibfnamefont {G.}~\bibnamefont
  {'t~Hooft}}\ and\ \bibinfo {author} {\bibfnamefont {M.~J.~G.}\ \bibnamefont
  {Veltman}},\ }\href@noop {} {\bibfield  {journal} {\bibinfo  {journal} {Ann.
  Inst. H. Poincare Phys. Theor.}\ }\textbf {\bibinfo {volume} {A20}},\
  \bibinfo {pages} {69} (\bibinfo {year} {1974})}\BibitemShut {NoStop}%
\bibitem [{\citenamefont {Deser}(2000)}]{Deser:1999mh}%
  \BibitemOpen
  \bibfield  {author} {\bibinfo {author} {\bibfnamefont {S.}~\bibnamefont
  {Deser}},\ }\bibfield  {booktitle} {\emph {\bibinfo {booktitle}
  {{Gravitation. Proceedings, International European Conference, 27th Session
  of the Journees Relativistes, Weimar, Germany, September 12-17, 1999}}},\
  }\href {\doibase
  10.1002/(SICI)1521-3889(200005)9:3/5<299::AID-ANDP299>3.0.CO;2-E} {\bibfield
  {journal} {\bibinfo  {journal} {Annalen Phys.}\ }\textbf {\bibinfo {volume}
  {9}},\ \bibinfo {pages} {299} (\bibinfo {year} {2000})},\ \Eprint
  {http://arxiv.org/abs/gr-qc/9911073} {arXiv:gr-qc/9911073 [gr-qc]}
  \BibitemShut {NoStop}%
\bibitem [{\citenamefont {Aghanim}\ \emph {et~al.}(2018)\citenamefont {Aghanim}
  \emph {et~al.}}]{Aghanim:2018eyx}%
  \BibitemOpen
  \bibfield  {author} {\bibinfo {author} {\bibfnamefont {N.}~\bibnamefont
  {Aghanim}} \emph {et~al.} (\bibinfo {collaboration} {Planck}),\ }\href@noop
  {} {\  (\bibinfo {year} {2018})},\ \Eprint {http://arxiv.org/abs/1807.06209}
  {arXiv:1807.06209 [astro-ph.CO]} \BibitemShut {NoStop}%
\bibitem [{\citenamefont {Clifton}\ \emph {et~al.}(2012)\citenamefont
  {Clifton}, \citenamefont {Ferreira}, \citenamefont {Padilla},\ and\
  \citenamefont {Skordis}}]{Clifton:2011jh}%
  \BibitemOpen
  \bibfield  {author} {\bibinfo {author} {\bibfnamefont {T.}~\bibnamefont
  {Clifton}}, \bibinfo {author} {\bibfnamefont {P.~G.}\ \bibnamefont
  {Ferreira}}, \bibinfo {author} {\bibfnamefont {A.}~\bibnamefont {Padilla}}, \
  and\ \bibinfo {author} {\bibfnamefont {C.}~\bibnamefont {Skordis}},\ }\href
  {\doibase 10.1016/j.physrep.2012.01.001} {\bibfield  {journal} {\bibinfo
  {journal} {Phys. Rept.}\ }\textbf {\bibinfo {volume} {513}},\ \bibinfo
  {pages} {1} (\bibinfo {year} {2012})},\ \Eprint
  {http://arxiv.org/abs/1106.2476} {arXiv:1106.2476 [astro-ph.CO]} \BibitemShut
  {NoStop}%
\bibitem [{\citenamefont {Koyama}(2016)}]{Koyama:2015vza}%
  \BibitemOpen
  \bibfield  {author} {\bibinfo {author} {\bibfnamefont {K.}~\bibnamefont
  {Koyama}},\ }\href {\doibase 10.1088/0034-4885/79/4/046902} {\bibfield
  {journal} {\bibinfo  {journal} {Rept. Prog. Phys.}\ }\textbf {\bibinfo
  {volume} {79}},\ \bibinfo {pages} {046902} (\bibinfo {year} {2016})},\
  \Eprint {http://arxiv.org/abs/1504.04623} {arXiv:1504.04623 [astro-ph.CO]}
  \BibitemShut {NoStop}%
\bibitem [{\citenamefont {Joyce}\ \emph {et~al.}(2015)\citenamefont {Joyce},
  \citenamefont {Jain}, \citenamefont {Khoury},\ and\ \citenamefont
  {Trodden}}]{Joyce:2014kja}%
  \BibitemOpen
  \bibfield  {author} {\bibinfo {author} {\bibfnamefont {A.}~\bibnamefont
  {Joyce}}, \bibinfo {author} {\bibfnamefont {B.}~\bibnamefont {Jain}},
  \bibinfo {author} {\bibfnamefont {J.}~\bibnamefont {Khoury}}, \ and\ \bibinfo
  {author} {\bibfnamefont {M.}~\bibnamefont {Trodden}},\ }\href {\doibase
  10.1016/j.physrep.2014.12.002} {\bibfield  {journal} {\bibinfo  {journal}
  {Phys. Rept.}\ }\textbf {\bibinfo {volume} {568}},\ \bibinfo {pages} {1}
  (\bibinfo {year} {2015})},\ \Eprint {http://arxiv.org/abs/1407.0059}
  {arXiv:1407.0059 [astro-ph.CO]} \BibitemShut {NoStop}%
\bibitem [{\citenamefont {Wei}(2018)}]{Wei:2018cov}%
  \BibitemOpen
  \bibfield  {author} {\bibinfo {author} {\bibfnamefont {J.-J.}\ \bibnamefont
  {Wei}},\ }\href@noop {} {\  (\bibinfo {year} {2018})},\ \Eprint
  {http://arxiv.org/abs/1806.09781} {arXiv:1806.09781 [astro-ph.CO]}
  \BibitemShut {NoStop}%
\bibitem [{\citenamefont {Wei}\ and\ \citenamefont {Wu}(2017)}]{Wei:2017emo}%
  \BibitemOpen
  \bibfield  {author} {\bibinfo {author} {\bibfnamefont {J.-J.}\ \bibnamefont
  {Wei}}\ and\ \bibinfo {author} {\bibfnamefont {X.-F.}\ \bibnamefont {Wu}},\
  }\href {\doibase 10.1093/mnras/stx2210} {\bibfield  {journal} {\bibinfo
  {journal} {Mon. Not. Roy. Astron. Soc.}\ }\textbf {\bibinfo {volume} {472}},\
  \bibinfo {pages} {2906} (\bibinfo {year} {2017})},\ \Eprint
  {http://arxiv.org/abs/1707.04152} {arXiv:1707.04152 [astro-ph.CO]}
  \BibitemShut {NoStop}%
\bibitem [{\citenamefont {Baker}\ and\ \citenamefont
  {Trodden}(2017)}]{Baker:2016reh}%
  \BibitemOpen
  \bibfield  {author} {\bibinfo {author} {\bibfnamefont {T.}~\bibnamefont
  {Baker}}\ and\ \bibinfo {author} {\bibfnamefont {M.}~\bibnamefont
  {Trodden}},\ }\href {\doibase 10.1103/PhysRevD.95.063512} {\bibfield
  {journal} {\bibinfo  {journal} {Phys. Rev.}\ }\textbf {\bibinfo {volume}
  {D95}},\ \bibinfo {pages} {063512} (\bibinfo {year} {2017})},\ \Eprint
  {http://arxiv.org/abs/1612.02004} {arXiv:1612.02004 [astro-ph.CO]}
  \BibitemShut {NoStop}%
\bibitem [{\citenamefont {Collett}\ and\ \citenamefont
  {Bacon}(2017)}]{Collett:2016dey}%
  \BibitemOpen
  \bibfield  {author} {\bibinfo {author} {\bibfnamefont {T.~E.}\ \bibnamefont
  {Collett}}\ and\ \bibinfo {author} {\bibfnamefont {D.}~\bibnamefont
  {Bacon}},\ }\href {\doibase 10.1103/PhysRevLett.118.091101} {\bibfield
  {journal} {\bibinfo  {journal} {Phys. Rev. Lett.}\ }\textbf {\bibinfo
  {volume} {118}},\ \bibinfo {pages} {091101} (\bibinfo {year} {2017})},\
  \Eprint {http://arxiv.org/abs/1602.05882} {arXiv:1602.05882 [astro-ph.HE]}
  \BibitemShut {NoStop}%
\bibitem [{\citenamefont {et.
  al.}(2017{\natexlab{a}})}]{PhysRevLett.119.161101}%
  \BibitemOpen
  \bibfield  {author} {\bibinfo {author} {\bibfnamefont {B.~P.~A.}\
  \bibnamefont {et. al.}} (\bibinfo {collaboration} {LIGO Scientific
  Collaboration and Virgo Collaboration}),\ }\href {\doibase
  10.1103/PhysRevLett.119.161101} {\bibfield  {journal} {\bibinfo  {journal}
  {Phys. Rev. Lett.}\ }\textbf {\bibinfo {volume} {119}},\ \bibinfo {pages}
  {161101} (\bibinfo {year} {2017}{\natexlab{a}})}\BibitemShut {NoStop}%
\bibitem [{\citenamefont {{LIGO Scientific Collaboration}}\ \emph
  {et~al.}(2015)\citenamefont {{LIGO Scientific Collaboration}}, \citenamefont
  {{Aasi}}, \citenamefont {{Abbott}}, \citenamefont {{Abbott}}, \citenamefont
  {{Abbott}}, \citenamefont {{Abernathy}}, \citenamefont {{Ackley}},
  \citenamefont {{Adams}}, \citenamefont {{Adams}}, \citenamefont {{Addesso}},\
  and\ \citenamefont {et~al.}}]{LIGO}%
  \BibitemOpen
  \bibfield  {author} {\bibinfo {author} {\bibnamefont {{LIGO Scientific
  Collaboration}}}, \bibinfo {author} {\bibfnamefont {J.}~\bibnamefont
  {{Aasi}}}, \bibinfo {author} {\bibfnamefont {B.~P.}\ \bibnamefont
  {{Abbott}}}, \bibinfo {author} {\bibfnamefont {R.}~\bibnamefont {{Abbott}}},
  \bibinfo {author} {\bibfnamefont {T.}~\bibnamefont {{Abbott}}}, \bibinfo
  {author} {\bibfnamefont {M.~R.}\ \bibnamefont {{Abernathy}}}, \bibinfo
  {author} {\bibfnamefont {K.}~\bibnamefont {{Ackley}}}, \bibinfo {author}
  {\bibfnamefont {C.}~\bibnamefont {{Adams}}}, \bibinfo {author} {\bibfnamefont
  {T.}~\bibnamefont {{Adams}}}, \bibinfo {author} {\bibfnamefont
  {P.}~\bibnamefont {{Addesso}}}, \ and\ \bibinfo {author} {\bibnamefont
  {et~al.}},\ }\href {\doibase 10.1088/0264-9381/32/7/074001} {\bibfield
  {journal} {\bibinfo  {journal} {Classical and Quantum Gravity}\ }\textbf
  {\bibinfo {volume} {32}},\ \bibinfo {eid} {074001} (\bibinfo {year}
  {2015})},\ \Eprint {http://arxiv.org/abs/1411.4547} {arXiv:1411.4547 [gr-qc]}
  \BibitemShut {NoStop}%
\bibitem [{\citenamefont {{Acernese}}\ \emph {et~al.}(2015)\citenamefont
  {{Acernese}}, \citenamefont {{Agathos}}, \citenamefont {{Agatsuma}},
  \citenamefont {{Aisa}}, \citenamefont {{Allemandou}}, \citenamefont
  {{Allocca}}, \citenamefont {{Amarni}}, \citenamefont {{Astone}},
  \citenamefont {{Balestri}}, \citenamefont {{Ballardin}},\ and\ \citenamefont
  {et~al.}}]{Virgo}%
  \BibitemOpen
  \bibfield  {author} {\bibinfo {author} {\bibfnamefont {F.}~\bibnamefont
  {{Acernese}}}, \bibinfo {author} {\bibfnamefont {M.}~\bibnamefont
  {{Agathos}}}, \bibinfo {author} {\bibfnamefont {K.}~\bibnamefont
  {{Agatsuma}}}, \bibinfo {author} {\bibfnamefont {D.}~\bibnamefont {{Aisa}}},
  \bibinfo {author} {\bibfnamefont {N.}~\bibnamefont {{Allemandou}}}, \bibinfo
  {author} {\bibfnamefont {A.}~\bibnamefont {{Allocca}}}, \bibinfo {author}
  {\bibfnamefont {J.}~\bibnamefont {{Amarni}}}, \bibinfo {author}
  {\bibfnamefont {P.}~\bibnamefont {{Astone}}}, \bibinfo {author}
  {\bibfnamefont {G.}~\bibnamefont {{Balestri}}}, \bibinfo {author}
  {\bibfnamefont {G.}~\bibnamefont {{Ballardin}}}, \ and\ \bibinfo {author}
  {\bibnamefont {et~al.}},\ }\href {\doibase 10.1088/0264-9381/32/2/024001}
  {\bibfield  {journal} {\bibinfo  {journal} {Classical and Quantum Gravity}\
  }\textbf {\bibinfo {volume} {32}},\ \bibinfo {eid} {024001} (\bibinfo {year}
  {2015})},\ \Eprint {http://arxiv.org/abs/1408.3978} {arXiv:1408.3978 [gr-qc]}
  \BibitemShut {NoStop}%
\bibitem [{\citenamefont {et. al.}(2017{\natexlab{b}})}]{2041-8205-848-2-L14}%
  \BibitemOpen
  \bibfield  {author} {\bibinfo {author} {\bibfnamefont {A.~G.}\ \bibnamefont
  {et. al.}},\ }\href {http://stacks.iop.org/2041-8205/848/i=2/a=L14}
  {\bibfield  {journal} {\bibinfo  {journal} {The Astrophysical Journal
  Letters}\ }\textbf {\bibinfo {volume} {848}},\ \bibinfo {pages} {L14}
  (\bibinfo {year} {2017}{\natexlab{b}})}\BibitemShut {NoStop}%
\bibitem [{\citenamefont {et. al.}(2017{\natexlab{c}})}]{2041-8205-848-2-L15}%
  \BibitemOpen
  \bibfield  {author} {\bibinfo {author} {\bibfnamefont {V.~S.}\ \bibnamefont
  {et. al.}},\ }\href {http://stacks.iop.org/2041-8205/848/i=2/a=L15}
  {\bibfield  {journal} {\bibinfo  {journal} {The Astrophysical Journal
  Letters}\ }\textbf {\bibinfo {volume} {848}},\ \bibinfo {pages} {L15}
  (\bibinfo {year} {2017}{\natexlab{c}})}\BibitemShut {NoStop}%
\bibitem [{\citenamefont {Baker}\ \emph {et~al.}(2017)\citenamefont {Baker},
  \citenamefont {Bellini}, \citenamefont {Ferreira}, \citenamefont {Lagos},
  \citenamefont {Noller},\ and\ \citenamefont {Sawicki}}]{Baker:2017hug}%
  \BibitemOpen
  \bibfield  {author} {\bibinfo {author} {\bibfnamefont {T.}~\bibnamefont
  {Baker}}, \bibinfo {author} {\bibfnamefont {E.}~\bibnamefont {Bellini}},
  \bibinfo {author} {\bibfnamefont {P.~G.}\ \bibnamefont {Ferreira}}, \bibinfo
  {author} {\bibfnamefont {M.}~\bibnamefont {Lagos}}, \bibinfo {author}
  {\bibfnamefont {J.}~\bibnamefont {Noller}}, \ and\ \bibinfo {author}
  {\bibfnamefont {I.}~\bibnamefont {Sawicki}},\ }\href@noop {} {\  (\bibinfo
  {year} {2017})},\ \Eprint {http://arxiv.org/abs/1710.06394} {arXiv:1710.06394
  [astro-ph.CO]} \BibitemShut {NoStop}%
\bibitem [{\citenamefont {Creminelli}\ and\ \citenamefont
  {Vernizzi}(2017)}]{Creminelli:2017sry}%
  \BibitemOpen
  \bibfield  {author} {\bibinfo {author} {\bibfnamefont {P.}~\bibnamefont
  {Creminelli}}\ and\ \bibinfo {author} {\bibfnamefont {F.}~\bibnamefont
  {Vernizzi}},\ }\href@noop {} {\  (\bibinfo {year} {2017})},\ \Eprint
  {http://arxiv.org/abs/1710.05877} {arXiv:1710.05877 [astro-ph.CO]}
  \BibitemShut {NoStop}%
\bibitem [{\citenamefont {Sakstein}\ and\ \citenamefont
  {Jain}(2017)}]{Sakstein:2017xjx}%
  \BibitemOpen
  \bibfield  {author} {\bibinfo {author} {\bibfnamefont {J.}~\bibnamefont
  {Sakstein}}\ and\ \bibinfo {author} {\bibfnamefont {B.}~\bibnamefont
  {Jain}},\ }\href@noop {} {\  (\bibinfo {year} {2017})},\ \Eprint
  {http://arxiv.org/abs/1710.05893} {arXiv:1710.05893 [astro-ph.CO]}
  \BibitemShut {NoStop}%
\bibitem [{\citenamefont {Ezquiaga}\ and\ \citenamefont
  {Zumalacarregui}(2017)}]{Ezquiaga:2017ekz}%
  \BibitemOpen
  \bibfield  {author} {\bibinfo {author} {\bibfnamefont {J.~M.}\ \bibnamefont
  {Ezquiaga}}\ and\ \bibinfo {author} {\bibfnamefont {M.}~\bibnamefont
  {Zumalacarregui}},\ }\href@noop {} {\  (\bibinfo {year} {2017})},\ \Eprint
  {http://arxiv.org/abs/1710.05901} {arXiv:1710.05901 [astro-ph.CO]}
  \BibitemShut {NoStop}%
\bibitem [{\citenamefont {Wang}\ \emph {et~al.}(2017)\citenamefont {Wang} \emph
  {et~al.}}]{Wang:2017rpx}%
  \BibitemOpen
  \bibfield  {author} {\bibinfo {author} {\bibfnamefont {H.}~\bibnamefont
  {Wang}} \emph {et~al.},\ }\href@noop {} {\  (\bibinfo {year} {2017})},\
  \Eprint {http://arxiv.org/abs/1710.05805} {arXiv:1710.05805 [astro-ph.HE]}
  \BibitemShut {NoStop}%
\bibitem [{\citenamefont {Lombriser}\ and\ \citenamefont
  {Lima}(2017)}]{Lombriser:2016yzn}%
  \BibitemOpen
  \bibfield  {author} {\bibinfo {author} {\bibfnamefont {L.}~\bibnamefont
  {Lombriser}}\ and\ \bibinfo {author} {\bibfnamefont {N.~A.}\ \bibnamefont
  {Lima}},\ }\href {\doibase 10.1016/j.physletb.2016.12.048} {\bibfield
  {journal} {\bibinfo  {journal} {Phys. Lett.}\ }\textbf {\bibinfo {volume}
  {B765}},\ \bibinfo {pages} {382} (\bibinfo {year} {2017})},\ \Eprint
  {http://arxiv.org/abs/1602.07670} {arXiv:1602.07670 [astro-ph.CO]}
  \BibitemShut {NoStop}%
\bibitem [{\citenamefont {Abbott}\ \emph {et~al.}(2017)\citenamefont {Abbott}
  \emph {et~al.}}]{Abbott:2017xzu}%
  \BibitemOpen
  \bibfield  {author} {\bibinfo {author} {\bibfnamefont {B.~P.}\ \bibnamefont
  {Abbott}} \emph {et~al.} (\bibinfo {collaboration} {LIGO Scientific,
  VINROUGE, Las Cumbres Observatory, DES, DLT40, Virgo, 1M2H, Dark Energy
  Camera GW-E, MASTER}),\ }\href {\doibase 10.1038/nature24471} {\bibfield
  {journal} {\bibinfo  {journal} {Nature}\ }\textbf {\bibinfo {volume} {551}},\
  \bibinfo {pages} {85} (\bibinfo {year} {2017})},\ \Eprint
  {http://arxiv.org/abs/1710.05835} {arXiv:1710.05835 [astro-ph.CO]}
  \BibitemShut {NoStop}%
\bibitem [{\citenamefont {{The Planck Collaboration}}(2006)}]{Planck}%
  \BibitemOpen
  \bibfield  {author} {\bibinfo {author} {\bibnamefont {{The Planck
  Collaboration}}},\ }\href@noop {} {\bibfield  {journal} {\bibinfo  {journal}
  {ArXiv Astrophysics e-prints}\ } (\bibinfo {year} {2006})},\ \Eprint
  {http://arxiv.org/abs/astro-ph/0604069} {astro-ph/0604069} \BibitemShut
  {NoStop}%
\bibitem [{\citenamefont {{Riess}}(2006)}]{SHoES}%
  \BibitemOpen
  \bibfield  {author} {\bibinfo {author} {\bibfnamefont {A.}~\bibnamefont
  {{Riess}}},\ }\href@noop {} {\enquote {\bibinfo {title} {{SHOES-Supernovae,
  HO, for the Equation of State of Dark energy}},}\ }\bibinfo {howpublished}
  {HST Proposal} (\bibinfo {year} {2006})\BibitemShut {NoStop}%
\bibitem [{\citenamefont {{Riess}}\ \emph {et~al.}(2018)\citenamefont
  {{Riess}}, \citenamefont {{Casertano}}, \citenamefont {{Yuan}}, \citenamefont
  {{Macri}}, \citenamefont {{Anderson}}, \citenamefont {{MacKenty}},
  \citenamefont {{Bowers}}, \citenamefont {{Clubb}}, \citenamefont
  {{Filippenko}}, \citenamefont {{Jones}},\ and\ \citenamefont
  {{Tucker}}}]{2018ApJ...855..136R}%
  \BibitemOpen
  \bibfield  {author} {\bibinfo {author} {\bibfnamefont {A.~G.}\ \bibnamefont
  {{Riess}}}, \bibinfo {author} {\bibfnamefont {S.}~\bibnamefont
  {{Casertano}}}, \bibinfo {author} {\bibfnamefont {W.}~\bibnamefont {{Yuan}}},
  \bibinfo {author} {\bibfnamefont {L.}~\bibnamefont {{Macri}}}, \bibinfo
  {author} {\bibfnamefont {J.}~\bibnamefont {{Anderson}}}, \bibinfo {author}
  {\bibfnamefont {J.~W.}\ \bibnamefont {{MacKenty}}}, \bibinfo {author}
  {\bibfnamefont {J.~B.}\ \bibnamefont {{Bowers}}}, \bibinfo {author}
  {\bibfnamefont {K.~I.}\ \bibnamefont {{Clubb}}}, \bibinfo {author}
  {\bibfnamefont {A.~V.}\ \bibnamefont {{Filippenko}}}, \bibinfo {author}
  {\bibfnamefont {D.~O.}\ \bibnamefont {{Jones}}}, \ and\ \bibinfo {author}
  {\bibfnamefont {B.~E.}\ \bibnamefont {{Tucker}}},\ }\href {\doibase
  10.3847/1538-4357/aaadb7} {\bibfield  {journal} {\bibinfo  {journal} {\apj}\
  }\textbf {\bibinfo {volume} {855}},\ \bibinfo {eid} {136} (\bibinfo {year}
  {2018})},\ \Eprint {http://arxiv.org/abs/1801.01120} {arXiv:1801.01120
  [astro-ph.SR]} \BibitemShut {NoStop}%
\bibitem [{\citenamefont {{Schutz}}(1986)}]{Schutz:1986}%
  \BibitemOpen
  \bibfield  {author} {\bibinfo {author} {\bibfnamefont {B.~F.}\ \bibnamefont
  {{Schutz}}},\ }\href {\doibase 10.1038/323310a0} {\bibfield  {journal}
  {\bibinfo  {journal} {Nature}\ }\textbf {\bibinfo {volume} {323}},\ \bibinfo
  {pages} {310} (\bibinfo {year} {1986})}\BibitemShut {NoStop}%
\bibitem [{\citenamefont {Holz}\ and\ \citenamefont
  {Hughes}(2005)}]{0004-637X-629-1-15}%
  \BibitemOpen
  \bibfield  {author} {\bibinfo {author} {\bibfnamefont {D.~E.}\ \bibnamefont
  {Holz}}\ and\ \bibinfo {author} {\bibfnamefont {S.~A.}\ \bibnamefont
  {Hughes}},\ }\href {http://stacks.iop.org/0004-637X/629/i=1/a=15} {\bibfield
  {journal} {\bibinfo  {journal} {The Astrophysical Journal}\ }\textbf
  {\bibinfo {volume} {629}},\ \bibinfo {pages} {15} (\bibinfo {year}
  {2005})}\BibitemShut {NoStop}%
\bibitem [{\citenamefont {Chen}\ \emph {et~al.}(2017)\citenamefont {Chen},
  \citenamefont {Fishbach},\ and\ \citenamefont {Holz}}]{Chen:2017rfc}%
  \BibitemOpen
  \bibfield  {author} {\bibinfo {author} {\bibfnamefont {H.-Y.}\ \bibnamefont
  {Chen}}, \bibinfo {author} {\bibfnamefont {M.}~\bibnamefont {Fishbach}}, \
  and\ \bibinfo {author} {\bibfnamefont {D.~E.}\ \bibnamefont {Holz}},\
  }\href@noop {} {\  (\bibinfo {year} {2017})},\ \Eprint
  {http://arxiv.org/abs/1712.06531} {arXiv:1712.06531 [astro-ph.CO]}
  \BibitemShut {NoStop}%
\bibitem [{\citenamefont {{Nissanke}}\ \emph {et~al.}(2013)\citenamefont
  {{Nissanke}}, \citenamefont {{Holz}}, \citenamefont {{Dalal}}, \citenamefont
  {{Hughes}}, \citenamefont {{Sievers}},\ and\ \citenamefont
  {{Hirata}}}]{Nissanke:2013}%
  \BibitemOpen
  \bibfield  {author} {\bibinfo {author} {\bibfnamefont {S.}~\bibnamefont
  {{Nissanke}}}, \bibinfo {author} {\bibfnamefont {D.~E.}\ \bibnamefont
  {{Holz}}}, \bibinfo {author} {\bibfnamefont {N.}~\bibnamefont {{Dalal}}},
  \bibinfo {author} {\bibfnamefont {S.~A.}\ \bibnamefont {{Hughes}}}, \bibinfo
  {author} {\bibfnamefont {J.~L.}\ \bibnamefont {{Sievers}}}, \ and\ \bibinfo
  {author} {\bibfnamefont {C.~M.}\ \bibnamefont {{Hirata}}},\ }\href@noop {}
  {\bibfield  {journal} {\bibinfo  {journal} {arXiv e-prints}\ ,\ \bibinfo
  {eid} {arXiv:1307.2638}} (\bibinfo {year} {2013})},\ \Eprint
  {http://arxiv.org/abs/1307.2638} {arXiv:1307.2638 [astro-ph.CO]} \BibitemShut
  {NoStop}%
\bibitem [{\citenamefont {Feeney}\ \emph {et~al.}(2018)\citenamefont {Feeney},
  \citenamefont {Peiris}, \citenamefont {Williamson}, \citenamefont {Nissanke},
  \citenamefont {Mortlock}, \citenamefont {Alsing},\ and\ \citenamefont
  {Scolnic}}]{Feeney:2018mkj}%
  \BibitemOpen
  \bibfield  {author} {\bibinfo {author} {\bibfnamefont {S.~M.}\ \bibnamefont
  {Feeney}}, \bibinfo {author} {\bibfnamefont {H.~V.}\ \bibnamefont {Peiris}},
  \bibinfo {author} {\bibfnamefont {A.~R.}\ \bibnamefont {Williamson}},
  \bibinfo {author} {\bibfnamefont {S.~M.}\ \bibnamefont {Nissanke}}, \bibinfo
  {author} {\bibfnamefont {D.~J.}\ \bibnamefont {Mortlock}}, \bibinfo {author}
  {\bibfnamefont {J.}~\bibnamefont {Alsing}}, \ and\ \bibinfo {author}
  {\bibfnamefont {D.}~\bibnamefont {Scolnic}},\ }\href@noop {} {\  (\bibinfo
  {year} {2018})},\ \Eprint {http://arxiv.org/abs/1802.03404} {arXiv:1802.03404
  [astro-ph.CO]} \BibitemShut {NoStop}%
\bibitem [{\citenamefont {{Aso}}\ \emph {et~al.}(2013)\citenamefont {{Aso}},
  \citenamefont {{Michimura}}, \citenamefont {{Somiya}}, \citenamefont
  {{Ando}}, \citenamefont {{Miyakawa}}, \citenamefont {{Sekiguchi}},
  \citenamefont {{Tatsumi}},\ and\ \citenamefont {{Yamamoto}}}]{KAGRA}%
  \BibitemOpen
  \bibfield  {author} {\bibinfo {author} {\bibfnamefont {Y.}~\bibnamefont
  {{Aso}}}, \bibinfo {author} {\bibfnamefont {Y.}~\bibnamefont {{Michimura}}},
  \bibinfo {author} {\bibfnamefont {K.}~\bibnamefont {{Somiya}}}, \bibinfo
  {author} {\bibfnamefont {M.}~\bibnamefont {{Ando}}}, \bibinfo {author}
  {\bibfnamefont {O.}~\bibnamefont {{Miyakawa}}}, \bibinfo {author}
  {\bibfnamefont {T.}~\bibnamefont {{Sekiguchi}}}, \bibinfo {author}
  {\bibfnamefont {D.}~\bibnamefont {{Tatsumi}}}, \ and\ \bibinfo {author}
  {\bibfnamefont {H.}~\bibnamefont {{Yamamoto}}},\ }\href {\doibase
  10.1103/PhysRevD.88.043007} {\bibfield  {journal} {\bibinfo  {journal}
  {\prd}\ }\textbf {\bibinfo {volume} {88}},\ \bibinfo {eid} {043007} (\bibinfo
  {year} {2013})},\ \Eprint {http://arxiv.org/abs/1306.6747} {arXiv:1306.6747
  [gr-qc]} \BibitemShut {NoStop}%
\bibitem [{\citenamefont {Iyer}\ \emph {et~al.}(2011)\citenamefont {Iyer},
  \citenamefont {Souradeep}, \citenamefont {Unnikrishnan}, \citenamefont
  {Dhurandhar}, \citenamefont {Raja},\ and\ \citenamefont {Sengupta}}]{IndIGO}%
  \BibitemOpen
  \bibfield  {author} {\bibinfo {author} {\bibfnamefont {B.}~\bibnamefont
  {Iyer}}, \bibinfo {author} {\bibfnamefont {T.}~\bibnamefont {Souradeep}},
  \bibinfo {author} {\bibfnamefont {C.~S.}\ \bibnamefont {Unnikrishnan}},
  \bibinfo {author} {\bibfnamefont {S.}~\bibnamefont {Dhurandhar}}, \bibinfo
  {author} {\bibfnamefont {S.}~\bibnamefont {Raja}}, \ and\ \bibinfo {author}
  {\bibfnamefont {A.}~\bibnamefont {Sengupta}},\ }\href
  {https://dcc.ligo.org/LIGO-M1100296/public} {\enquote {\bibinfo {title}
  {{LIGO-India, Proposal of the Consortium for Indian Initiative in
  Gravitational-wave Observations (IndIGO)}},}\ } (\bibinfo {year}
  {2011})\BibitemShut {NoStop}%
\bibitem [{\citenamefont {McClelland}\ \emph {et~al.}(2015)\citenamefont
  {McClelland}, \citenamefont {Evans}, \citenamefont {Lantz}, \citenamefont
  {Martin}, \citenamefont {Quetschke},\ and\ \citenamefont {Schnabel}}]{Aplus}%
  \BibitemOpen
  \bibfield  {author} {\bibinfo {author} {\bibfnamefont {D.}~\bibnamefont
  {McClelland}}, \bibinfo {author} {\bibfnamefont {M.}~\bibnamefont {Evans}},
  \bibinfo {author} {\bibfnamefont {B.}~\bibnamefont {Lantz}}, \bibinfo
  {author} {\bibfnamefont {I.}~\bibnamefont {Martin}}, \bibinfo {author}
  {\bibfnamefont {V.}~\bibnamefont {Quetschke}}, \ and\ \bibinfo {author}
  {\bibfnamefont {R.}~\bibnamefont {Schnabel}},\ }\href
  {https://dcc.ligo.org/LIGO-T1500290/public} {\enquote {\bibinfo {title}
  {{Instrument Science White Paper 2015}},}\ } (\bibinfo {year}
  {2015})\BibitemShut {NoStop}%
\bibitem [{\citenamefont {{The LIGO Scientific Collaboration}}\ and\
  \citenamefont {{the Virgo Collaboration}}(2018)}]{Catalog}%
  \BibitemOpen
  \bibfield  {author} {\bibinfo {author} {\bibnamefont {{The LIGO Scientific
  Collaboration}}}\ and\ \bibinfo {author} {\bibnamefont {{the Virgo
  Collaboration}}},\ }\href@noop {} {\bibfield  {journal} {\bibinfo  {journal}
  {arXiv e-prints}\ ,\ \bibinfo {eid} {arXiv:1811.12907}} (\bibinfo {year}
  {2018})},\ \Eprint {http://arxiv.org/abs/1811.12907} {arXiv:1811.12907
  [astro-ph.HE]} \BibitemShut {NoStop}%
\bibitem [{\citenamefont {Ezquiaga}\ and\ \citenamefont
  {Zumalacarregui}(2018)}]{Ezquiaga:2018btd}%
  \BibitemOpen
  \bibfield  {author} {\bibinfo {author} {\bibfnamefont {J.~M.}\ \bibnamefont
  {Ezquiaga}}\ and\ \bibinfo {author} {\bibfnamefont {M.}~\bibnamefont
  {Zumalacarregui}},\ }\href@noop {} {\  (\bibinfo {year} {2018})},\ \Eprint
  {http://arxiv.org/abs/1807.09241} {arXiv:1807.09241 [astro-ph.CO]}
  \BibitemShut {NoStop}%
\bibitem [{\citenamefont {Belgacem}\ \emph {et~al.}(2018)\citenamefont
  {Belgacem}, \citenamefont {Dirian}, \citenamefont {Foffa},\ and\
  \citenamefont {Maggiore}}]{Belgacem:2018lbp}%
  \BibitemOpen
  \bibfield  {author} {\bibinfo {author} {\bibfnamefont {E.}~\bibnamefont
  {Belgacem}}, \bibinfo {author} {\bibfnamefont {Y.}~\bibnamefont {Dirian}},
  \bibinfo {author} {\bibfnamefont {S.}~\bibnamefont {Foffa}}, \ and\ \bibinfo
  {author} {\bibfnamefont {M.}~\bibnamefont {Maggiore}},\ }\href@noop {} {\
  (\bibinfo {year} {2018})},\ \Eprint {http://arxiv.org/abs/1805.08731}
  {arXiv:1805.08731 [gr-qc]} \BibitemShut {NoStop}%
\bibitem [{\citenamefont {Uzan}(2011)}]{Uzan:2010pm}%
  \BibitemOpen
  \bibfield  {author} {\bibinfo {author} {\bibfnamefont {J.-P.}\ \bibnamefont
  {Uzan}},\ }\href {\doibase 10.12942/lrr-2011-2} {\bibfield  {journal}
  {\bibinfo  {journal} {Living Rev. Rel.}\ }\textbf {\bibinfo {volume} {14}},\
  \bibinfo {pages} {2} (\bibinfo {year} {2011})},\ \Eprint
  {http://arxiv.org/abs/1009.5514} {arXiv:1009.5514 [astro-ph.CO]} \BibitemShut
  {NoStop}%
\bibitem [{\citenamefont {Pettorino}\ and\ \citenamefont
  {Amendola}(2015)}]{Pettorino:2014bka}%
  \BibitemOpen
  \bibfield  {author} {\bibinfo {author} {\bibfnamefont {V.}~\bibnamefont
  {Pettorino}}\ and\ \bibinfo {author} {\bibfnamefont {L.}~\bibnamefont
  {Amendola}},\ }\href {\doibase 10.1016/j.physletb.2015.02.007} {\bibfield
  {journal} {\bibinfo  {journal} {Phys. Lett.}\ }\textbf {\bibinfo {volume}
  {B742}},\ \bibinfo {pages} {353} (\bibinfo {year} {2015})},\ \Eprint
  {http://arxiv.org/abs/1408.2224} {arXiv:1408.2224 [astro-ph.CO]} \BibitemShut
  {NoStop}%
\bibitem [{\citenamefont {Lombriser}\ and\ \citenamefont
  {Taylor}(2016)}]{Lombriser:2015sxa}%
  \BibitemOpen
  \bibfield  {author} {\bibinfo {author} {\bibfnamefont {L.}~\bibnamefont
  {Lombriser}}\ and\ \bibinfo {author} {\bibfnamefont {A.}~\bibnamefont
  {Taylor}},\ }\href {\doibase 10.1088/1475-7516/2016/03/031} {\bibfield
  {journal} {\bibinfo  {journal} {JCAP}\ }\textbf {\bibinfo {volume} {1603}},\
  \bibinfo {pages} {031} (\bibinfo {year} {2016})},\ \Eprint
  {http://arxiv.org/abs/1509.08458} {arXiv:1509.08458 [astro-ph.CO]}
  \BibitemShut {NoStop}%
\bibitem [{\citenamefont {Nishizawa}(2018)}]{Nishizawa:2017nef}%
  \BibitemOpen
  \bibfield  {author} {\bibinfo {author} {\bibfnamefont {A.}~\bibnamefont
  {Nishizawa}},\ }\href {\doibase 10.1103/PhysRevD.97.104037} {\bibfield
  {journal} {\bibinfo  {journal} {Phys. Rev.}\ }\textbf {\bibinfo {volume}
  {D97}},\ \bibinfo {pages} {104037} (\bibinfo {year} {2018})},\ \Eprint
  {http://arxiv.org/abs/1710.04825} {arXiv:1710.04825 [gr-qc]} \BibitemShut
  {NoStop}%
\bibitem [{\citenamefont {Arai}\ and\ \citenamefont
  {Nishizawa}(2018)}]{Arai:2017hxj}%
  \BibitemOpen
  \bibfield  {author} {\bibinfo {author} {\bibfnamefont {S.}~\bibnamefont
  {Arai}}\ and\ \bibinfo {author} {\bibfnamefont {A.}~\bibnamefont
  {Nishizawa}},\ }\href {\doibase 10.1103/PhysRevD.97.104038} {\bibfield
  {journal} {\bibinfo  {journal} {Phys. Rev.}\ }\textbf {\bibinfo {volume}
  {D97}},\ \bibinfo {pages} {104038} (\bibinfo {year} {2018})},\ \Eprint
  {http://arxiv.org/abs/1711.03776} {arXiv:1711.03776 [gr-qc]} \BibitemShut
  {NoStop}%
\bibitem [{\citenamefont {Amendola}\ \emph {et~al.}(2017)\citenamefont
  {Amendola}, \citenamefont {Sawicki}, \citenamefont {Kunz},\ and\
  \citenamefont {Saltas}}]{Amendola:2017ovw}%
  \BibitemOpen
  \bibfield  {author} {\bibinfo {author} {\bibfnamefont {L.}~\bibnamefont
  {Amendola}}, \bibinfo {author} {\bibfnamefont {I.}~\bibnamefont {Sawicki}},
  \bibinfo {author} {\bibfnamefont {M.}~\bibnamefont {Kunz}}, \ and\ \bibinfo
  {author} {\bibfnamefont {I.~D.}\ \bibnamefont {Saltas}},\ }\href@noop {} {\
  (\bibinfo {year} {2017})},\ \Eprint {http://arxiv.org/abs/1712.08623}
  {arXiv:1712.08623 [astro-ph.CO]} \BibitemShut {NoStop}%
\bibitem [{\citenamefont {Linder}(2018)}]{Linder:2018jil}%
  \BibitemOpen
  \bibfield  {author} {\bibinfo {author} {\bibfnamefont {E.~V.}\ \bibnamefont
  {Linder}},\ }\href {\doibase 10.1088/1475-7516/2018/03/005} {\bibfield
  {journal} {\bibinfo  {journal} {JCAP}\ }\textbf {\bibinfo {volume} {1803}},\
  \bibinfo {pages} {005} (\bibinfo {year} {2018})},\ \Eprint
  {http://arxiv.org/abs/1801.01503} {arXiv:1801.01503 [astro-ph.CO]}
  \BibitemShut {NoStop}%
\bibitem [{\citenamefont {Noller}\ and\ \citenamefont
  {Nicola}(2018)}]{Noller:2018wyv}%
  \BibitemOpen
  \bibfield  {author} {\bibinfo {author} {\bibfnamefont {J.}~\bibnamefont
  {Noller}}\ and\ \bibinfo {author} {\bibfnamefont {A.}~\bibnamefont
  {Nicola}},\ }\href@noop {} {\  (\bibinfo {year} {2018})},\ \Eprint
  {http://arxiv.org/abs/1811.12928} {arXiv:1811.12928 [astro-ph.CO]}
  \BibitemShut {NoStop}%
\bibitem [{\citenamefont {de~Rham}\ \emph {et~al.}(2011)\citenamefont
  {de~Rham}, \citenamefont {Gabadadze},\ and\ \citenamefont
  {Tolley}}]{deRham:2010kj}%
  \BibitemOpen
  \bibfield  {author} {\bibinfo {author} {\bibfnamefont {C.}~\bibnamefont
  {de~Rham}}, \bibinfo {author} {\bibfnamefont {G.}~\bibnamefont {Gabadadze}},
  \ and\ \bibinfo {author} {\bibfnamefont {A.~J.}\ \bibnamefont {Tolley}},\
  }\href {\doibase 10.1103/PhysRevLett.106.231101} {\bibfield  {journal}
  {\bibinfo  {journal} {Phys.Rev.Lett.}\ }\textbf {\bibinfo {volume} {106}},\
  \bibinfo {pages} {231101} (\bibinfo {year} {2011})},\ \Eprint
  {http://arxiv.org/abs/1011.1232} {arXiv:1011.1232 [hep-th]} \BibitemShut
  {NoStop}%
\bibitem [{\citenamefont {de~Rham}\ and\ \citenamefont
  {Gabadadze}(2010)}]{deRham:2010ik}%
  \BibitemOpen
  \bibfield  {author} {\bibinfo {author} {\bibfnamefont {C.}~\bibnamefont
  {de~Rham}}\ and\ \bibinfo {author} {\bibfnamefont {G.}~\bibnamefont
  {Gabadadze}},\ }\href {\doibase 10.1103/PhysRevD.82.044020} {\bibfield
  {journal} {\bibinfo  {journal} {Phys.Rev.}\ }\textbf {\bibinfo {volume}
  {D82}},\ \bibinfo {pages} {044020} (\bibinfo {year} {2010})},\ \Eprint
  {http://arxiv.org/abs/1007.0443} {arXiv:1007.0443 [hep-th]} \BibitemShut
  {NoStop}%
\bibitem [{\citenamefont {Hassan}\ and\ \citenamefont
  {Rosen}(2012)}]{Hassan:2011zd}%
  \BibitemOpen
  \bibfield  {author} {\bibinfo {author} {\bibfnamefont {S.}~\bibnamefont
  {Hassan}}\ and\ \bibinfo {author} {\bibfnamefont {R.~A.}\ \bibnamefont
  {Rosen}},\ }\href {\doibase 10.1007/JHEP02(2012)126} {\bibfield  {journal}
  {\bibinfo  {journal} {JHEP}\ }\textbf {\bibinfo {volume} {1202}},\ \bibinfo
  {pages} {126} (\bibinfo {year} {2012})},\ \Eprint
  {http://arxiv.org/abs/1109.3515} {arXiv:1109.3515 [hep-th]} \BibitemShut
  {NoStop}%
\bibitem [{\citenamefont {Max}\ \emph {et~al.}(2017)\citenamefont {Max},
  \citenamefont {Platscher},\ and\ \citenamefont {Smirnov}}]{Max:2017flc}%
  \BibitemOpen
  \bibfield  {author} {\bibinfo {author} {\bibfnamefont {K.}~\bibnamefont
  {Max}}, \bibinfo {author} {\bibfnamefont {M.}~\bibnamefont {Platscher}}, \
  and\ \bibinfo {author} {\bibfnamefont {J.}~\bibnamefont {Smirnov}},\ }\href
  {\doibase 10.1103/PhysRevLett.119.111101} {\bibfield  {journal} {\bibinfo
  {journal} {Phys. Rev. Lett.}\ }\textbf {\bibinfo {volume} {119}},\ \bibinfo
  {pages} {111101} (\bibinfo {year} {2017})},\ \Eprint
  {http://arxiv.org/abs/1703.07785} {arXiv:1703.07785 [gr-qc]} \BibitemShut
  {NoStop}%
\bibitem [{\citenamefont {Max}\ \emph {et~al.}(2018)\citenamefont {Max},
  \citenamefont {Platscher},\ and\ \citenamefont {Smirnov}}]{Max:2017kdc}%
  \BibitemOpen
  \bibfield  {author} {\bibinfo {author} {\bibfnamefont {K.}~\bibnamefont
  {Max}}, \bibinfo {author} {\bibfnamefont {M.}~\bibnamefont {Platscher}}, \
  and\ \bibinfo {author} {\bibfnamefont {J.}~\bibnamefont {Smirnov}},\ }\href
  {\doibase 10.1103/PhysRevD.97.064009} {\bibfield  {journal} {\bibinfo
  {journal} {Phys. Rev.}\ }\textbf {\bibinfo {volume} {D97}},\ \bibinfo {pages}
  {064009} (\bibinfo {year} {2018})},\ \Eprint
  {http://arxiv.org/abs/1712.06601} {arXiv:1712.06601 [gr-qc]} \BibitemShut
  {NoStop}%
\bibitem [{\citenamefont {Beltran~Jimenez}\ and\ \citenamefont
  {Heisenberg}(2018)}]{BeltranJimenez:2018ymu}%
  \BibitemOpen
  \bibfield  {author} {\bibinfo {author} {\bibfnamefont {J.}~\bibnamefont
  {Beltran~Jimenez}}\ and\ \bibinfo {author} {\bibfnamefont {L.}~\bibnamefont
  {Heisenberg}},\ }\href@noop {} {\  (\bibinfo {year} {2018})},\ \Eprint
  {http://arxiv.org/abs/1806.01753} {arXiv:1806.01753 [gr-qc]} \BibitemShut
  {NoStop}%
\bibitem [{\citenamefont {Horndeski}(1974)}]{Horndeski:1974wa}%
  \BibitemOpen
  \bibfield  {author} {\bibinfo {author} {\bibfnamefont {G.~W.}\ \bibnamefont
  {Horndeski}},\ }\href {\doibase 10.1007/BF01807638} {\bibfield  {journal}
  {\bibinfo  {journal} {Int. J. Theor. Phys.}\ }\textbf {\bibinfo {volume}
  {10}},\ \bibinfo {pages} {363} (\bibinfo {year} {1974})}\BibitemShut
  {NoStop}%
\bibitem [{\citenamefont {Deffayet}\ \emph {et~al.}(2011)\citenamefont
  {Deffayet}, \citenamefont {Gao}, \citenamefont {Steer},\ and\ \citenamefont
  {Zahariade}}]{Deffayet:2011gz}%
  \BibitemOpen
  \bibfield  {author} {\bibinfo {author} {\bibfnamefont {C.}~\bibnamefont
  {Deffayet}}, \bibinfo {author} {\bibfnamefont {X.}~\bibnamefont {Gao}},
  \bibinfo {author} {\bibfnamefont {D.~A.}\ \bibnamefont {Steer}}, \ and\
  \bibinfo {author} {\bibfnamefont {G.}~\bibnamefont {Zahariade}},\ }\href
  {\doibase 10.1103/PhysRevD.84.064039} {\bibfield  {journal} {\bibinfo
  {journal} {Phys. Rev.}\ }\textbf {\bibinfo {volume} {D84}},\ \bibinfo {pages}
  {064039} (\bibinfo {year} {2011})},\ \Eprint {http://arxiv.org/abs/1103.3260}
  {arXiv:1103.3260 [hep-th]} \BibitemShut {NoStop}%
\bibitem [{\citenamefont {Gleyzes}\ \emph {et~al.}(2015)\citenamefont
  {Gleyzes}, \citenamefont {Langlois}, \citenamefont {Piazza},\ and\
  \citenamefont {Vernizzi}}]{Gleyzes:2014dya}%
  \BibitemOpen
  \bibfield  {author} {\bibinfo {author} {\bibfnamefont {J.}~\bibnamefont
  {Gleyzes}}, \bibinfo {author} {\bibfnamefont {D.}~\bibnamefont {Langlois}},
  \bibinfo {author} {\bibfnamefont {F.}~\bibnamefont {Piazza}}, \ and\ \bibinfo
  {author} {\bibfnamefont {F.}~\bibnamefont {Vernizzi}},\ }\href {\doibase
  10.1103/PhysRevLett.114.211101} {\bibfield  {journal} {\bibinfo  {journal}
  {Phys. Rev. Lett.}\ }\textbf {\bibinfo {volume} {114}},\ \bibinfo {pages}
  {211101} (\bibinfo {year} {2015})},\ \Eprint {http://arxiv.org/abs/1404.6495}
  {arXiv:1404.6495 [hep-th]} \BibitemShut {NoStop}%
\bibitem [{\citenamefont {Zumalacarregui}\ and\ \citenamefont
  {Garcia-Bellido}(2014)}]{Zumalacarregui:2013pma}%
  \BibitemOpen
  \bibfield  {author} {\bibinfo {author} {\bibfnamefont {M.}~\bibnamefont
  {Zumalacarregui}}\ and\ \bibinfo {author} {\bibfnamefont {J.}~\bibnamefont
  {Garcia-Bellido}},\ }\href {\doibase 10.1103/PhysRevD.89.064046} {\bibfield
  {journal} {\bibinfo  {journal} {Phys. Rev.}\ }\textbf {\bibinfo {volume}
  {D89}},\ \bibinfo {pages} {064046} (\bibinfo {year} {2014})},\ \Eprint
  {http://arxiv.org/abs/1308.4685} {arXiv:1308.4685 [gr-qc]} \BibitemShut
  {NoStop}%
\bibitem [{\citenamefont {Langlois}\ and\ \citenamefont
  {Noui}(2016)}]{Langlois:2015cwa}%
  \BibitemOpen
  \bibfield  {author} {\bibinfo {author} {\bibfnamefont {D.}~\bibnamefont
  {Langlois}}\ and\ \bibinfo {author} {\bibfnamefont {K.}~\bibnamefont
  {Noui}},\ }\href {\doibase 10.1088/1475-7516/2016/02/034} {\bibfield
  {journal} {\bibinfo  {journal} {JCAP}\ }\textbf {\bibinfo {volume} {1602}},\
  \bibinfo {pages} {034} (\bibinfo {year} {2016})},\ \Eprint
  {http://arxiv.org/abs/1510.06930} {arXiv:1510.06930 [gr-qc]} \BibitemShut
  {NoStop}%
\bibitem [{\citenamefont {Crisostomi}\ \emph {et~al.}(2016)\citenamefont
  {Crisostomi}, \citenamefont {Koyama},\ and\ \citenamefont
  {Tasinato}}]{Crisostomi:2016czh}%
  \BibitemOpen
  \bibfield  {author} {\bibinfo {author} {\bibfnamefont {M.}~\bibnamefont
  {Crisostomi}}, \bibinfo {author} {\bibfnamefont {K.}~\bibnamefont {Koyama}},
  \ and\ \bibinfo {author} {\bibfnamefont {G.}~\bibnamefont {Tasinato}},\
  }\href {\doibase 10.1088/1475-7516/2016/04/044} {\bibfield  {journal}
  {\bibinfo  {journal} {JCAP}\ }\textbf {\bibinfo {volume} {1604}},\ \bibinfo
  {pages} {044} (\bibinfo {year} {2016})},\ \Eprint
  {http://arxiv.org/abs/1602.03119} {arXiv:1602.03119 [hep-th]} \BibitemShut
  {NoStop}%
\bibitem [{\citenamefont {Achour}\ \emph {et~al.}(2016)\citenamefont {Achour},
  \citenamefont {Langlois},\ and\ \citenamefont {Noui}}]{Achour:2016rkg}%
  \BibitemOpen
  \bibfield  {author} {\bibinfo {author} {\bibfnamefont {J.~B.}\ \bibnamefont
  {Achour}}, \bibinfo {author} {\bibfnamefont {D.}~\bibnamefont {Langlois}}, \
  and\ \bibinfo {author} {\bibfnamefont {K.}~\bibnamefont {Noui}},\ }\href@noop
  {} {\  (\bibinfo {year} {2016})},\ \Eprint {http://arxiv.org/abs/1602.08398}
  {arXiv:1602.08398 [gr-qc]} \BibitemShut {NoStop}%
\bibitem [{\citenamefont {Kase}\ and\ \citenamefont
  {Tsujikawa}(2018)}]{Kase:2018aps}%
  \BibitemOpen
  \bibfield  {author} {\bibinfo {author} {\bibfnamefont {R.}~\bibnamefont
  {Kase}}\ and\ \bibinfo {author} {\bibfnamefont {S.}~\bibnamefont
  {Tsujikawa}},\ }\href@noop {} {\  (\bibinfo {year} {2018})},\ \Eprint
  {http://arxiv.org/abs/1809.08735} {arXiv:1809.08735 [gr-qc]} \BibitemShut
  {NoStop}%
\bibitem [{\citenamefont {Langlois}(2018)}]{Langlois:2018dxi}%
  \BibitemOpen
  \bibfield  {author} {\bibinfo {author} {\bibfnamefont {D.}~\bibnamefont
  {Langlois}},\ }\href@noop {} {\  (\bibinfo {year} {2018})},\ \Eprint
  {http://arxiv.org/abs/1811.06271} {arXiv:1811.06271 [gr-qc]} \BibitemShut
  {NoStop}%
\bibitem [{\citenamefont {Kobayashi}(2019)}]{Kobayashi:2019hrl}%
  \BibitemOpen
  \bibfield  {author} {\bibinfo {author} {\bibfnamefont {T.}~\bibnamefont
  {Kobayashi}},\ }\href@noop {} {\  (\bibinfo {year} {2019})},\ \Eprint
  {http://arxiv.org/abs/1901.07183} {arXiv:1901.07183 [gr-qc]} \BibitemShut
  {NoStop}%
\bibitem [{\citenamefont {Heisenberg}(2014)}]{Heisenberg:2014rta}%
  \BibitemOpen
  \bibfield  {author} {\bibinfo {author} {\bibfnamefont {L.}~\bibnamefont
  {Heisenberg}},\ }\href {\doibase 10.1088/1475-7516/2014/05/015} {\bibfield
  {journal} {\bibinfo  {journal} {JCAP}\ }\textbf {\bibinfo {volume} {1405}},\
  \bibinfo {pages} {015} (\bibinfo {year} {2014})},\ \Eprint
  {http://arxiv.org/abs/1402.7026} {arXiv:1402.7026 [hep-th]} \BibitemShut
  {NoStop}%
\bibitem [{\citenamefont {Jacobson}\ and\ \citenamefont
  {Mattingly}(2001)}]{Jacobson:2000xp}%
  \BibitemOpen
  \bibfield  {author} {\bibinfo {author} {\bibfnamefont {T.}~\bibnamefont
  {Jacobson}}\ and\ \bibinfo {author} {\bibfnamefont {D.}~\bibnamefont
  {Mattingly}},\ }\href {\doibase 10.1103/PhysRevD.64.024028} {\bibfield
  {journal} {\bibinfo  {journal} {Phys. Rev.}\ }\textbf {\bibinfo {volume}
  {D64}},\ \bibinfo {pages} {024028} (\bibinfo {year} {2001})},\ \Eprint
  {http://arxiv.org/abs/gr-qc/0007031} {arXiv:gr-qc/0007031 [gr-qc]}
  \BibitemShut {NoStop}%
\bibitem [{\citenamefont {Zlosnik}\ \emph {et~al.}(2006)\citenamefont
  {Zlosnik}, \citenamefont {Ferreira},\ and\ \citenamefont
  {Starkman}}]{Zlosnik:2006sb}%
  \BibitemOpen
  \bibfield  {author} {\bibinfo {author} {\bibfnamefont {T.~G.}\ \bibnamefont
  {Zlosnik}}, \bibinfo {author} {\bibfnamefont {P.~G.}\ \bibnamefont
  {Ferreira}}, \ and\ \bibinfo {author} {\bibfnamefont {G.~D.}\ \bibnamefont
  {Starkman}},\ }\href {\doibase 10.1103/PhysRevD.74.044037} {\bibfield
  {journal} {\bibinfo  {journal} {Phys. Rev.}\ }\textbf {\bibinfo {volume}
  {D74}},\ \bibinfo {pages} {044037} (\bibinfo {year} {2006})},\ \Eprint
  {http://arxiv.org/abs/gr-qc/0606039} {arXiv:gr-qc/0606039 [gr-qc]}
  \BibitemShut {NoStop}%
\bibitem [{\citenamefont {Heisenberg}\ \emph {et~al.}(2018)\citenamefont
  {Heisenberg}, \citenamefont {Kase},\ and\ \citenamefont
  {Tsujikawa}}]{Heisenberg:2018mxx}%
  \BibitemOpen
  \bibfield  {author} {\bibinfo {author} {\bibfnamefont {L.}~\bibnamefont
  {Heisenberg}}, \bibinfo {author} {\bibfnamefont {R.}~\bibnamefont {Kase}}, \
  and\ \bibinfo {author} {\bibfnamefont {S.}~\bibnamefont {Tsujikawa}},\ }\href
  {\doibase 10.1103/PhysRevD.98.024038} {\bibfield  {journal} {\bibinfo
  {journal} {Phys. Rev.}\ }\textbf {\bibinfo {volume} {D98}},\ \bibinfo {pages}
  {024038} (\bibinfo {year} {2018})},\ \Eprint
  {http://arxiv.org/abs/1805.01066} {arXiv:1805.01066 [gr-qc]} \BibitemShut
  {NoStop}%
\bibitem [{\citenamefont {Kreisch}\ and\ \citenamefont
  {Komatsu}(2018)}]{Kreisch:2017uet}%
  \BibitemOpen
  \bibfield  {author} {\bibinfo {author} {\bibfnamefont {C.~D.}\ \bibnamefont
  {Kreisch}}\ and\ \bibinfo {author} {\bibfnamefont {E.}~\bibnamefont
  {Komatsu}},\ }\href {\doibase 10.1088/1475-7516/2018/12/030} {\bibfield
  {journal} {\bibinfo  {journal} {JCAP}\ }\textbf {\bibinfo {volume} {1812}},\
  \bibinfo {pages} {030} (\bibinfo {year} {2018})},\ \Eprint
  {http://arxiv.org/abs/1712.02710} {arXiv:1712.02710 [astro-ph.CO]}
  \BibitemShut {NoStop}%
\bibitem [{\citenamefont {Ade}\ \emph {et~al.}(2015)\citenamefont {Ade} \emph
  {et~al.}}]{Ade:2014zfo}%
  \BibitemOpen
  \bibfield  {author} {\bibinfo {author} {\bibfnamefont {P.~A.~R.}\
  \bibnamefont {Ade}} \emph {et~al.} (\bibinfo {collaboration} {Planck}),\
  }\href {\doibase 10.1051/0004-6361/201424496} {\bibfield  {journal} {\bibinfo
   {journal} {Astron. Astrophys.}\ }\textbf {\bibinfo {volume} {580}},\
  \bibinfo {pages} {A22} (\bibinfo {year} {2015})},\ \Eprint
  {http://arxiv.org/abs/1406.7482} {arXiv:1406.7482 [astro-ph.CO]} \BibitemShut
  {NoStop}%
\bibitem [{\citenamefont {Ade}\ \emph {et~al.}(2016)\citenamefont {Ade} \emph
  {et~al.}}]{Ade:2015rim}%
  \BibitemOpen
  \bibfield  {author} {\bibinfo {author} {\bibfnamefont {P.~A.~R.}\
  \bibnamefont {Ade}} \emph {et~al.} (\bibinfo {collaboration} {Planck}),\
  }\href {\doibase 10.1051/0004-6361/201525814} {\bibfield  {journal} {\bibinfo
   {journal} {Astron. Astrophys.}\ }\textbf {\bibinfo {volume} {594}},\
  \bibinfo {pages} {A14} (\bibinfo {year} {2016})},\ \Eprint
  {http://arxiv.org/abs/1502.01590} {arXiv:1502.01590 [astro-ph.CO]}
  \BibitemShut {NoStop}%
\bibitem [{\citenamefont {Huang}(2016)}]{Huang:2015srv}%
  \BibitemOpen
  \bibfield  {author} {\bibinfo {author} {\bibfnamefont {Z.}~\bibnamefont
  {Huang}},\ }\href {\doibase 10.1103/PhysRevD.93.043538} {\bibfield  {journal}
  {\bibinfo  {journal} {Phys. Rev.}\ }\textbf {\bibinfo {volume} {D93}},\
  \bibinfo {pages} {043538} (\bibinfo {year} {2016})},\ \Eprint
  {http://arxiv.org/abs/1511.02808} {arXiv:1511.02808 [astro-ph.CO]}
  \BibitemShut {NoStop}%
\bibitem [{\citenamefont {Copi}\ \emph {et~al.}(2004)\citenamefont {Copi},
  \citenamefont {Davis},\ and\ \citenamefont {Krauss}}]{Copi:2003xd}%
  \BibitemOpen
  \bibfield  {author} {\bibinfo {author} {\bibfnamefont {C.~J.}\ \bibnamefont
  {Copi}}, \bibinfo {author} {\bibfnamefont {A.~N.}\ \bibnamefont {Davis}}, \
  and\ \bibinfo {author} {\bibfnamefont {L.~M.}\ \bibnamefont {Krauss}},\
  }\href {\doibase 10.1103/PhysRevLett.92.171301} {\bibfield  {journal}
  {\bibinfo  {journal} {Phys. Rev. Lett.}\ }\textbf {\bibinfo {volume} {92}},\
  \bibinfo {pages} {171301} (\bibinfo {year} {2004})},\ \Eprint
  {http://arxiv.org/abs/astro-ph/0311334} {arXiv:astro-ph/0311334 [astro-ph]}
  \BibitemShut {NoStop}%
\bibitem [{\citenamefont {Bellini}\ and\ \citenamefont
  {Sawicki}(2014)}]{Bellini:2014fua}%
  \BibitemOpen
  \bibfield  {author} {\bibinfo {author} {\bibfnamefont {E.}~\bibnamefont
  {Bellini}}\ and\ \bibinfo {author} {\bibfnamefont {I.}~\bibnamefont
  {Sawicki}},\ }\href {\doibase 10.1088/1475-7516/2014/07/050} {\bibfield
  {journal} {\bibinfo  {journal} {JCAP}\ }\textbf {\bibinfo {volume} {1407}},\
  \bibinfo {pages} {050} (\bibinfo {year} {2014})},\ \Eprint
  {http://arxiv.org/abs/1404.3713} {arXiv:1404.3713 [astro-ph.CO]} \BibitemShut
  {NoStop}%
\bibitem [{\citenamefont {Lagos}\ \emph {et~al.}(2018)\citenamefont {Lagos},
  \citenamefont {Bellini}, \citenamefont {Noller}, \citenamefont {Ferreira},\
  and\ \citenamefont {Baker}}]{Lagos:2017hdr}%
  \BibitemOpen
  \bibfield  {author} {\bibinfo {author} {\bibfnamefont {M.}~\bibnamefont
  {Lagos}}, \bibinfo {author} {\bibfnamefont {E.}~\bibnamefont {Bellini}},
  \bibinfo {author} {\bibfnamefont {J.}~\bibnamefont {Noller}}, \bibinfo
  {author} {\bibfnamefont {P.~G.}\ \bibnamefont {Ferreira}}, \ and\ \bibinfo
  {author} {\bibfnamefont {T.}~\bibnamefont {Baker}},\ }\href {\doibase
  10.1088/1475-7516/2018/03/021} {\bibfield  {journal} {\bibinfo  {journal}
  {JCAP}\ }\textbf {\bibinfo {volume} {1803}},\ \bibinfo {pages} {021}
  (\bibinfo {year} {2018})},\ \Eprint {http://arxiv.org/abs/1711.09893}
  {arXiv:1711.09893 [gr-qc]} \BibitemShut {NoStop}%
\bibitem [{\citenamefont {Bellini}\ \emph {et~al.}(2016)\citenamefont
  {Bellini}, \citenamefont {Cuesta}, \citenamefont {Jimenez},\ and\
  \citenamefont {Verde}}]{Bellini:2015xja}%
  \BibitemOpen
  \bibfield  {author} {\bibinfo {author} {\bibfnamefont {E.}~\bibnamefont
  {Bellini}}, \bibinfo {author} {\bibfnamefont {A.~J.}\ \bibnamefont {Cuesta}},
  \bibinfo {author} {\bibfnamefont {R.}~\bibnamefont {Jimenez}}, \ and\
  \bibinfo {author} {\bibfnamefont {L.}~\bibnamefont {Verde}},\ }\href
  {\doibase 10.1088/1475-7516/2016/06/E01, 10.1088/1475-7516/2016/02/053}
  {\bibfield  {journal} {\bibinfo  {journal} {JCAP}\ }\textbf {\bibinfo
  {volume} {1602}},\ \bibinfo {pages} {053} (\bibinfo {year} {2016})},\
  \bibinfo {note} {[Erratum: JCAP1606,no.06,E01(2016)]},\ \Eprint
  {http://arxiv.org/abs/1509.07816} {arXiv:1509.07816 [astro-ph.CO]}
  \BibitemShut {NoStop}%
\bibitem [{\citenamefont {Williams}\ \emph {et~al.}(2004)\citenamefont
  {Williams}, \citenamefont {Turyshev},\ and\ \citenamefont
  {Boggs}}]{Williams:2004qba}%
  \BibitemOpen
  \bibfield  {author} {\bibinfo {author} {\bibfnamefont {J.~G.}\ \bibnamefont
  {Williams}}, \bibinfo {author} {\bibfnamefont {S.~G.}\ \bibnamefont
  {Turyshev}}, \ and\ \bibinfo {author} {\bibfnamefont {D.~H.}\ \bibnamefont
  {Boggs}},\ }\href {\doibase 10.1103/PhysRevLett.93.261101} {\bibfield
  {journal} {\bibinfo  {journal} {Phys. Rev. Lett.}\ }\textbf {\bibinfo
  {volume} {93}},\ \bibinfo {pages} {261101} (\bibinfo {year} {2004})},\
  \Eprint {http://arxiv.org/abs/gr-qc/0411113} {arXiv:gr-qc/0411113 [gr-qc]}
  \BibitemShut {NoStop}%
\bibitem [{\citenamefont {Hellings}\ \emph {et~al.}(1983)\citenamefont
  {Hellings}, \citenamefont {Adams}, \citenamefont {Anderson}, \citenamefont
  {Keesey}, \citenamefont {Lau}, \citenamefont {Standish}, \citenamefont
  {Canuto},\ and\ \citenamefont {Goldman}}]{PhysRevLett.51.1609}%
  \BibitemOpen
  \bibfield  {author} {\bibinfo {author} {\bibfnamefont {R.~W.}\ \bibnamefont
  {Hellings}}, \bibinfo {author} {\bibfnamefont {P.~J.}\ \bibnamefont {Adams}},
  \bibinfo {author} {\bibfnamefont {J.~D.}\ \bibnamefont {Anderson}}, \bibinfo
  {author} {\bibfnamefont {M.~S.}\ \bibnamefont {Keesey}}, \bibinfo {author}
  {\bibfnamefont {E.~L.}\ \bibnamefont {Lau}}, \bibinfo {author} {\bibfnamefont
  {E.~M.}\ \bibnamefont {Standish}}, \bibinfo {author} {\bibfnamefont {V.~M.}\
  \bibnamefont {Canuto}}, \ and\ \bibinfo {author} {\bibfnamefont
  {I.}~\bibnamefont {Goldman}},\ }\href {\doibase 10.1103/PhysRevLett.51.1609}
  {\bibfield  {journal} {\bibinfo  {journal} {Phys. Rev. Lett.}\ }\textbf
  {\bibinfo {volume} {51}},\ \bibinfo {pages} {1609} (\bibinfo {year}
  {1983})}\BibitemShut {NoStop}%
\bibitem [{\citenamefont {Bonanno}\ and\ \citenamefont
  {Froehlich}(2017)}]{Bonanno:2017dcx}%
  \BibitemOpen
  \bibfield  {author} {\bibinfo {author} {\bibfnamefont {A.}~\bibnamefont
  {Bonanno}}\ and\ \bibinfo {author} {\bibfnamefont {H.-E.}\ \bibnamefont
  {Froehlich}},\ }\href@noop {} {\  (\bibinfo {year} {2017})},\ \Eprint
  {http://arxiv.org/abs/1707.01866} {arXiv:1707.01866 [astro-ph.SR]}
  \BibitemShut {NoStop}%
\bibitem [{\citenamefont {Jain}\ and\ \citenamefont
  {Khoury}(2010)}]{Jain:2010ka}%
  \BibitemOpen
  \bibfield  {author} {\bibinfo {author} {\bibfnamefont {B.}~\bibnamefont
  {Jain}}\ and\ \bibinfo {author} {\bibfnamefont {J.}~\bibnamefont {Khoury}},\
  }\href {\doibase 10.1016/j.aop.2010.04.002} {\bibfield  {journal} {\bibinfo
  {journal} {Annals Phys.}\ }\textbf {\bibinfo {volume} {325}},\ \bibinfo
  {pages} {1479} (\bibinfo {year} {2010})},\ \Eprint
  {http://arxiv.org/abs/1004.3294} {arXiv:1004.3294 [astro-ph.CO]} \BibitemShut
  {NoStop}%
\bibitem [{\citenamefont {Tattersall}\ \emph {et~al.}(2018)\citenamefont
  {Tattersall}, \citenamefont {Ferreira},\ and\ \citenamefont
  {Lagos}}]{Tattersall:2017erk}%
  \BibitemOpen
  \bibfield  {author} {\bibinfo {author} {\bibfnamefont {O.~J.}\ \bibnamefont
  {Tattersall}}, \bibinfo {author} {\bibfnamefont {P.~G.}\ \bibnamefont
  {Ferreira}}, \ and\ \bibinfo {author} {\bibfnamefont {M.}~\bibnamefont
  {Lagos}},\ }\href {\doibase 10.1103/PhysRevD.97.044021} {\bibfield  {journal}
  {\bibinfo  {journal} {Phys. Rev.}\ }\textbf {\bibinfo {volume} {D97}},\
  \bibinfo {pages} {044021} (\bibinfo {year} {2018})},\ \Eprint
  {http://arxiv.org/abs/1711.01992} {arXiv:1711.01992 [gr-qc]} \BibitemShut
  {NoStop}%
\bibitem [{\citenamefont {Alonso}\ \emph {et~al.}(2017)\citenamefont {Alonso},
  \citenamefont {Bellini}, \citenamefont {Ferreira},\ and\ \citenamefont
  {Zumalacarregui}}]{Alonso:2016suf}%
  \BibitemOpen
  \bibfield  {author} {\bibinfo {author} {\bibfnamefont {D.}~\bibnamefont
  {Alonso}}, \bibinfo {author} {\bibfnamefont {E.}~\bibnamefont {Bellini}},
  \bibinfo {author} {\bibfnamefont {P.~G.}\ \bibnamefont {Ferreira}}, \ and\
  \bibinfo {author} {\bibfnamefont {M.}~\bibnamefont {Zumalacarregui}},\ }\href
  {\doibase 10.1103/PhysRevD.95.063502} {\bibfield  {journal} {\bibinfo
  {journal} {Phys. Rev.}\ }\textbf {\bibinfo {volume} {D95}},\ \bibinfo {pages}
  {063502} (\bibinfo {year} {2017})},\ \Eprint
  {http://arxiv.org/abs/1610.09290} {arXiv:1610.09290 [astro-ph.CO]}
  \BibitemShut {NoStop}%
\bibitem [{\citenamefont {Linder}(2017)}]{Linder:2016wqw}%
  \BibitemOpen
  \bibfield  {author} {\bibinfo {author} {\bibfnamefont {E.~V.}\ \bibnamefont
  {Linder}},\ }\href {\doibase 10.1103/PhysRevD.95.023518} {\bibfield
  {journal} {\bibinfo  {journal} {Phys. Rev.}\ }\textbf {\bibinfo {volume}
  {D95}},\ \bibinfo {pages} {023518} (\bibinfo {year} {2017})},\ \Eprint
  {http://arxiv.org/abs/1607.03113} {arXiv:1607.03113 [astro-ph.CO]}
  \BibitemShut {NoStop}%
\bibitem [{\citenamefont {Gleyzes}(2017)}]{Gleyzes:2017kpi}%
  \BibitemOpen
  \bibfield  {author} {\bibinfo {author} {\bibfnamefont {J.}~\bibnamefont
  {Gleyzes}},\ }\href {\doibase 10.1103/PhysRevD.96.063516} {\bibfield
  {journal} {\bibinfo  {journal} {Phys. Rev.}\ }\textbf {\bibinfo {volume}
  {D96}},\ \bibinfo {pages} {063516} (\bibinfo {year} {2017})},\ \Eprint
  {http://arxiv.org/abs/1705.04714} {arXiv:1705.04714 [astro-ph.CO]}
  \BibitemShut {NoStop}%
\bibitem [{\citenamefont {Abell}\ \emph {et~al.}(2009)\citenamefont {Abell}
  \emph {et~al.}}]{Abell:2009aa}%
  \BibitemOpen
  \bibfield  {author} {\bibinfo {author} {\bibfnamefont {P.~A.}\ \bibnamefont
  {Abell}} \emph {et~al.} (\bibinfo {collaboration} {LSST Science, LSST
  Project}),\ }\href@noop {} {\  (\bibinfo {year} {2009})},\ \Eprint
  {http://arxiv.org/abs/0912.0201} {arXiv:0912.0201 [astro-ph.IM]} \BibitemShut
  {NoStop}%
\bibitem [{\citenamefont {Carilli}\ and\ \citenamefont
  {Rawlings}(2004)}]{Carilli:2004nx}%
  \BibitemOpen
  \bibfield  {author} {\bibinfo {author} {\bibfnamefont {C.~L.}\ \bibnamefont
  {Carilli}}\ and\ \bibinfo {author} {\bibfnamefont {S.}~\bibnamefont
  {Rawlings}},\ }\bibfield  {booktitle} {\emph {\bibinfo {booktitle}
  {{International SKA Conference 2003 Geraldton, Australia, July 27-August 2,
  2003}}},\ }\href {\doibase 10.1016/j.newar.2004.09.001} {\bibfield  {journal}
  {\bibinfo  {journal} {New Astron. Rev.}\ }\textbf {\bibinfo {volume} {48}},\
  \bibinfo {pages} {979} (\bibinfo {year} {2004})},\ \Eprint
  {http://arxiv.org/abs/astro-ph/0409274} {arXiv:astro-ph/0409274 [astro-ph]}
  \BibitemShut {NoStop}%
\bibitem [{\citenamefont {Abazajian}\ \emph {et~al.}(2016)\citenamefont
  {Abazajian} \emph {et~al.}}]{Abazajian:2016yjj}%
  \BibitemOpen
  \bibfield  {author} {\bibinfo {author} {\bibfnamefont {K.~N.}\ \bibnamefont
  {Abazajian}} \emph {et~al.} (\bibinfo {collaboration} {CMB-S4}),\ }\href@noop
  {} {\  (\bibinfo {year} {2016})},\ \Eprint {http://arxiv.org/abs/1610.02743}
  {arXiv:1610.02743 [astro-ph.CO]} \BibitemShut {NoStop}%
\bibitem [{\citenamefont {{Cutler}}\ and\ \citenamefont
  {{Flanagan}}(1994)}]{1994PhRvD..49.2658C}%
  \BibitemOpen
  \bibfield  {author} {\bibinfo {author} {\bibfnamefont {C.}~\bibnamefont
  {{Cutler}}}\ and\ \bibinfo {author} {\bibfnamefont {{\'E}.~E.}\ \bibnamefont
  {{Flanagan}}},\ }\href {\doibase 10.1103/PhysRevD.49.2658} {\bibfield
  {journal} {\bibinfo  {journal} {\prd}\ }\textbf {\bibinfo {volume} {49}},\
  \bibinfo {pages} {2658} (\bibinfo {year} {1994})},\ \Eprint
  {http://arxiv.org/abs/gr-qc/9402014} {arXiv:gr-qc/9402014 [gr-qc]}
  \BibitemShut {NoStop}%
\bibitem [{\citenamefont {{Singer}}\ and\ \citenamefont
  {{Price}}(2016)}]{2016PhRvD..93b4013S}%
  \BibitemOpen
  \bibfield  {author} {\bibinfo {author} {\bibfnamefont {L.~P.}\ \bibnamefont
  {{Singer}}}\ and\ \bibinfo {author} {\bibfnamefont {L.~R.}\ \bibnamefont
  {{Price}}},\ }\href {\doibase 10.1103/PhysRevD.93.024013} {\bibfield
  {journal} {\bibinfo  {journal} {\prd}\ }\textbf {\bibinfo {volume} {93}},\
  \bibinfo {eid} {024013} (\bibinfo {year} {2016})},\ \Eprint
  {http://arxiv.org/abs/1508.03634} {arXiv:1508.03634 [gr-qc]} \BibitemShut
  {NoStop}%
\bibitem [{\citenamefont {{Chen}}\ and\ \citenamefont
  {{Holz}}(2017)}]{2017ApJ...840...88C}%
  \BibitemOpen
  \bibfield  {author} {\bibinfo {author} {\bibfnamefont {H.-Y.}\ \bibnamefont
  {{Chen}}}\ and\ \bibinfo {author} {\bibfnamefont {D.~E.}\ \bibnamefont
  {{Holz}}},\ }\href {\doibase 10.3847/1538-4357/aa6f0d} {\bibfield  {journal}
  {\bibinfo  {journal} {\apj}\ }\textbf {\bibinfo {volume} {840}},\ \bibinfo
  {eid} {88} (\bibinfo {year} {2017})},\ \Eprint
  {http://arxiv.org/abs/1509.00055} {arXiv:1509.00055 [astro-ph.IM]}
  \BibitemShut {NoStop}%
\bibitem [{\citenamefont {{Wu}}\ and\ \citenamefont
  {{Huterer}}(2017)}]{2017MNRAS.471.4946W}%
  \BibitemOpen
  \bibfield  {author} {\bibinfo {author} {\bibfnamefont {H.-Y.}\ \bibnamefont
  {{Wu}}}\ and\ \bibinfo {author} {\bibfnamefont {D.}~\bibnamefont
  {{Huterer}}},\ }\href {\doibase 10.1093/mnras/stx1967} {\bibfield  {journal}
  {\bibinfo  {journal} {Mon. Not. Roy. Astron. Soc.}\ }\textbf {\bibinfo
  {volume} {471}},\ \bibinfo {pages} {4946} (\bibinfo {year} {2017})},\ \Eprint
  {http://arxiv.org/abs/1706.09723} {arXiv:1706.09723} \BibitemShut {NoStop}%
\bibitem [{\citenamefont {{Abbott}}\ \emph {et~al.}(2018)\citenamefont
  {{Abbott}}, \citenamefont {{Abbott}}, \citenamefont {{Abbott}}, \citenamefont
  {{Acernese}},\ and\ \citenamefont {{Ackley}}}]{2018arXiv180511579T}%
  \BibitemOpen
  \bibfield  {author} {\bibinfo {author} {\bibfnamefont {B.~P.}\ \bibnamefont
  {{Abbott}}}, \bibinfo {author} {\bibfnamefont {R.}~\bibnamefont {{Abbott}}},
  \bibinfo {author} {\bibfnamefont {T.~D.}\ \bibnamefont {{Abbott}}}, \bibinfo
  {author} {\bibfnamefont {F.}~\bibnamefont {{Acernese}}}, \ and\ \bibinfo
  {author} {\bibfnamefont {K.}~\bibnamefont {{Ackley}}},\ }\href@noop {}
  {\bibfield  {journal} {\bibinfo  {journal} {arXiv e-prints}\ ,\ \bibinfo
  {eid} {arXiv:1805.11579}} (\bibinfo {year} {2018})},\ \Eprint
  {http://arxiv.org/abs/1805.11579} {arXiv:1805.11579 [gr-qc]} \BibitemShut
  {NoStop}%
\bibitem [{\citenamefont {{Pardo}}\ \emph {et~al.}(2018)\citenamefont
  {{Pardo}}, \citenamefont {{Fishbach}}, \citenamefont {{Holz}},\ and\
  \citenamefont {{Spergel}}}]{Pardo:2018}%
  \BibitemOpen
  \bibfield  {author} {\bibinfo {author} {\bibfnamefont {K.}~\bibnamefont
  {{Pardo}}}, \bibinfo {author} {\bibfnamefont {M.}~\bibnamefont {{Fishbach}}},
  \bibinfo {author} {\bibfnamefont {D.~E.}\ \bibnamefont {{Holz}}}, \ and\
  \bibinfo {author} {\bibfnamefont {D.~N.}\ \bibnamefont {{Spergel}}},\ }\href
  {\doibase 10.1088/1475-7516/2018/07/048} {\bibfield  {journal} {\bibinfo
  {journal} {Journal of Cosmology and Astro-Particle Physics}\ }\textbf
  {\bibinfo {volume} {2018}},\ \bibinfo {eid} {048} (\bibinfo {year} {2018})},\
  \Eprint {http://arxiv.org/abs/1801.08160} {arXiv:1801.08160 [gr-qc]}
  \BibitemShut {NoStop}%
\bibitem [{\citenamefont {{Mandel}}\ \emph {et~al.}(2018)\citenamefont
  {{Mandel}}, \citenamefont {{Farr}},\ and\ \citenamefont
  {{Gair}}}]{Mandel:2018}%
  \BibitemOpen
  \bibfield  {author} {\bibinfo {author} {\bibfnamefont {I.}~\bibnamefont
  {{Mandel}}}, \bibinfo {author} {\bibfnamefont {W.~M.}\ \bibnamefont
  {{Farr}}}, \ and\ \bibinfo {author} {\bibfnamefont {J.~R.}\ \bibnamefont
  {{Gair}}},\ }\href@noop {} {\bibfield  {journal} {\bibinfo  {journal} {arXiv
  e-prints}\ ,\ \bibinfo {eid} {arXiv:1809.02063}} (\bibinfo {year} {2018})},\
  \Eprint {http://arxiv.org/abs/1809.02063} {arXiv:1809.02063
  [physics.data-an]} \BibitemShut {NoStop}%
\bibitem [{\citenamefont {{The LIGO Scientific Collaboration}}\ and\
  \citenamefont {{The Virgo Collaboration}}(2018)}]{O2pop}%
  \BibitemOpen
  \bibfield  {author} {\bibinfo {author} {\bibnamefont {{The LIGO Scientific
  Collaboration}}}\ and\ \bibinfo {author} {\bibnamefont {{The Virgo
  Collaboration}}},\ }\href@noop {} {\bibfield  {journal} {\bibinfo  {journal}
  {arXiv e-prints}\ ,\ \bibinfo {eid} {arXiv:1811.12940}} (\bibinfo {year}
  {2018})},\ \Eprint {http://arxiv.org/abs/1811.12940} {arXiv:1811.12940
  [astro-ph.HE]} \BibitemShut {NoStop}%
\bibitem [{\citenamefont {{Madau}}\ and\ \citenamefont
  {{Dickinson}}(2014)}]{2014ARA&A..52..415M}%
  \BibitemOpen
  \bibfield  {author} {\bibinfo {author} {\bibfnamefont {P.}~\bibnamefont
  {{Madau}}}\ and\ \bibinfo {author} {\bibfnamefont {M.}~\bibnamefont
  {{Dickinson}}},\ }\href {\doibase 10.1146/annurev-astro-081811-125615}
  {\bibfield  {journal} {\bibinfo  {journal} {Annual Review of Astronomy and
  Astrophysics}\ }\textbf {\bibinfo {volume} {52}},\ \bibinfo {pages} {415}
  (\bibinfo {year} {2014})},\ \Eprint {http://arxiv.org/abs/1403.0007}
  {arXiv:1403.0007 [astro-ph.CO]} \BibitemShut {NoStop}%
\end{thebibliography}%

 \end{document}